\documentclass[twoside,twocolumn,english,prb]{revtex4}
\usepackage[T1]{fontenc}
\usepackage[latin9]{inputenc}
\usepackage{color}
\usepackage{units}
\usepackage{amsmath}
\usepackage{graphicx}
\usepackage{amssymb}

\makeatletter

\providecommand{\tabularnewline}{\\}

\@ifundefined{textcolor}{}
{%
 \definecolor{BLACK}{gray}{0}
 \definecolor{WHITE}{gray}{1}
 \definecolor{RED}{rgb}{1,0,0}
 \definecolor{GREEN}{rgb}{0,1,0}
 \definecolor{BLUE}{rgb}{0,0,1}
 \definecolor{CYAN}{cmyk}{1,0,0,0}
 \definecolor{MAGENTA}{cmyk}{0,1,0,0}
 \definecolor{YELLOW}{cmyk}{0,0,1,0}
 }

\@ifundefined{definecolor}
 {\usepackage{color}}{}
\makeatother

\makeatother

\usepackage{babel}

\begin{document}

\title{Doping and temperature dependence of Mn 3\textit{d} states in A-site
ordered manganites}

\author{M. García-Fernández$^{1,2}$, U. Staub$^{1}$, Y. Bodenthin$^{1}$,
V. Pomjakushin$^{3}$, A. Mirone$^{4}$, J. Fernández-Rodríguez$^{4}$,
V. Scagnoli$^{4}$, A. M. Mulders$^{5,6,7}$, S. M. Lawrence$^{6}$
and E. Pomjakushina$^{8}$.}

\affiliation{\textit{\footnotesize $^{1}$ Swiss Light Source, Paul Scherrer Institut,
5232 Villigen PSI, Switzerland}}

\affiliation{$^{2}$ \textit{\footnotesize Département de Physique, Université
de Fribourg, CH-1700 Fribourg, Switzerland}}

\affiliation{\textit{\footnotesize $^{3}$ Laboratory for Neutron Scattering,
Paul Scherrer Institut \& ETH Zürich, 5232 Villigen PSI, Switzerland }}

\affiliation{\textit{\footnotesize $^{4}$ European Synchrotron Radiation Facility,
BP 220, 38043 Grenoble Cedex 9, France}}

\affiliation{\textit{\footnotesize $^{5}$School of Physical, Environmental and
Mathematical Sciences, UNSW@ADFA, Canberra ACT 2600, Australia }}

\affiliation{\textit{\footnotesize $^{6}$Department of Imaging and Applied Physics,
Curtin University of Technology, Perth, WA 6845, Australia }}

\affiliation{\textit{\footnotesize $^{7}$The Bragg Institute, Australian Nuclear
Science and Technology Organisation, Lucas Heights, NSW 2234, Australia}}

\affiliation{\textit{\footnotesize $^{8}$Laboratory for Developement and Methods,
Paul Scherrer Institut, 5232 Villigen PSI, Switzerland}}
\begin{abstract}
We present a systematic study of the electronic structure in A-site
ordered manganites as function of doping and temperature. The energy
dependencies observed with soft x-ray resonant diffraction (SXRD)
at the Mn L$_{2,3}$ edges are compared with structural investigations
using neutron powder diffraction as well as with cluster calculations.
The crystal structures obtained with neutron powder diffraction reflect
the various orbital and charge ordered phases and show an increase
of the Mn-O-Mn bond angle as function of doping and temperature. Cluster
calculations show that the observed spectral changes in SXRD as a
function of doping are more pronounced than expected from an increase
in bandwitdh due to the increase in Mn-O-Mn bond angle, and are best
described by holes that are distributed at the neighbouring oxygen
ions. These holes are not directly added to the Mn 3\textit{d} shell,
but centered at the Mn site. In contrast, the spectral changes in
SXRD as function of temperature are best described by an increase
of magnetic correlations. This demonstrates the strong correlations
between orbitals and magnetic moments of the 3\textit{d} states.
\end{abstract}
\maketitle

\section{Introduction}

Manganites have attracted a lot of attention in the past two decades
because they show very rich phase diagrams with interesting electronic
and magnetic properties that make them challenging to be described
from first principles. For manganites crystallized in the perovskite
structure ABO$_{3}$, the A-site of the perovskite is coordinated
by 12 oxygen sites with A as a trivalent ion $R^{3+}$ ($R\equiv$Lanthanide).
This A-site can be doped with a divalent cation $T^{2+}$, normally
Ca$^{2+}$, Ba$^{2+}$ or Sr$^{2+}$. This doping causes the Mn ions,
that occupy the B-site, to change its average electronic states from
Mn$^{3+}$ to Mn$^{4+}$. The B sites are sixfold coordinated and
the surrounding O$^{2-}$ ions form octahedral cages. Arising from
the strong coupling between the electric, magnetic and structural
properties present in these systems, the physical and structural properties
depend strongly on the doping content and on the nature of the A-site
cation. The most important of these properties are the appearance
of the colossal magneto resistance (CMR) with the occurrence of phase
separation and unusual spin, charge, lattice and orbital order.

\begin{figure}
\begin{centering}
\includegraphics[scale=0.4]{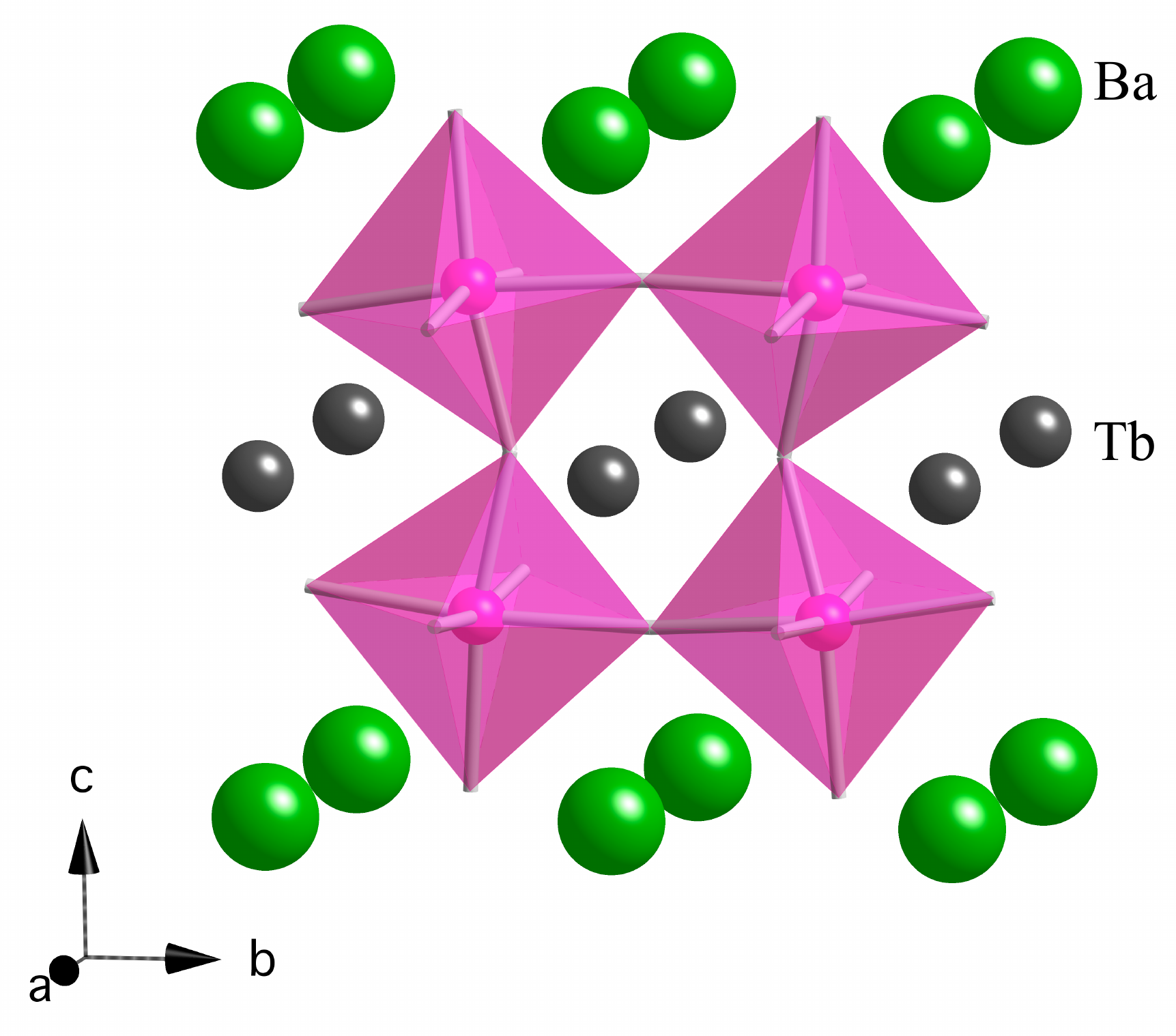}
\par\end{centering}

\caption{Perovskite structure of the A-site ordered half doped manganite TbBaMn$_{2}$O$_{6}$.
The A-site of the perovskite is occupied by Tb and Ba ordered in different
planes. The Mn ions are placed at the B-site within the oxygen octahedra.}

\end{figure}

For the half doped manganite, the generally assumed ground state consists
of the checkerboard CO (charge order) and OO (orbital order) pattern,
which is characterized by the alternation of Mn$^{3+}$ and Mn$^{4+}$
sites (Mn-centered charge ordering). In the last years other phases
have been proposed to be present in manganites close to half doping
\citep{Efremov2004}. Some of those phases present either bond-centered
charge ordering, allowing the existence of magnetic Zener polaron-type
phases; or an intermediate phase combining site centered CO and bond
centered CO that allows the occurrence of ferroelectricity. Experimentally,
Zener polaron phases have also been proposed for A-site ordered systems
of $R$BaMn$_{2}$O$_{6}$, in particular for $R=$Y. \citep{Daoud-Aladine2008}
However, recent resonant x-ray diffraction experiments found no inversion
symmetry breaking effects on the Mn sites confirming merely the checkerboard
charge and orbital ordering pattern for different doping contents
and materials studied \citep{Garcia-Fernandez2009}. The A-site randomness
(the $R^{3+}/T^{2+}$ solid solution) has a significant influence
on the properties of the system. It increases the magnetoresistance
effect and decreases the charge and orbital order temperature T$_{CO/OO}$.
It also makes the charge-spin-orbital correlation short-ranged \citep{Akahoshi2003,Akahoshi2006,Zimmermann1999}.
It is therefore important to study the A-site ordered manganite system
to quantify the effect of the quenched disorder on the CMR physics.

The crystal structure of the cation-ordered material $R$BaMn$_{2}$O$_{6}$
is of $a_{p}\mathrm{x}a_{p}\mathrm{x}2a_{p}$ type with $a_{p}\approx3.9\textrm{\AA}$
being the cubic perovskite unit cell parameter (see Figure 1). It
was shown that the A-site ordered half doped system SmBaMn$_{2}$O$_{6}$
exhibits a CO/OO transition at T$_{CO/OO}\approx360$$\:$K and an
antiferromagnetic transition at T$_{N}=250\: K$ followed by a reorientation
of the OO at T$_{CO2}=200\: K$ \citep{Akahoshi2006,Garcia-Fernandez-Sm-2008}.
Resonant soft x-ray diffraction has found an OO of e$_{g}$ electrons
with $\nicefrac{\left[x^{2}-z^{2}\right]}{\left[y^{2}-z^{2}\right]}$
type \citep{Garcia-Fernandez-Sm-2008}. When the system is hole doped,
a linear dependence of the ordering wave vector of the orbital reflection
$\overrightarrow{k}=\left(\delta_{x},\delta_{x},0\right)$ with doping
\textit{x} was found with $\delta=\frac{\left(1-x\right)}{4}$ in
Tb$_{1-x}$Ca$_{x}$BaMn$_{2}$O$_{6}$, while, $\overrightarrow{k}$
was found to be constant $\overrightarrow{k}=\left(\delta_{y},\delta_{y},0\right)=\left(\nicefrac{1}{4},\nicefrac{1}{4},0\right)$
for doping contents equal and below half doping TbBa$_{1-y}$La$_{y}$Mn$_{2}$O$_{6}$
\citep{Akahoshi2006,Garcia-Fernandez2009}. One possible OO layout
for the $\nicefrac{2}{3}$ doped system is shown in Figure 2.

Our previous resonant x-ray diffraction study showed that the spectral
shape of the orbital reflection exhibits distinct changes as function
of doping at the Mn $L_{2,3}$ edges \citep{Garcia-Fernandez2009}.
The characteristic features of the spectra are similar as for other
manganites \citep{Thomas2004,Staub2005,Dhesi2004,Wilkins2003,Beale2009,Staub2009},
but the relative intensities are altered and indicate a progressive
alteration of the electronic Mn states as a function of hole doping.
It was deduced that the anomalous OO melting and change in $\left(\delta_{x},\:\delta_{x},\:0\right)$
with hole doping are not related to the structure nor to the magnetic
interactions, but rather due to increased two dimensional character
of the orbital interactions. %This indicates an alteration of the electronic Mn states as a function of doping. 

\begin{figure}
\begin{centering}
\includegraphics[scale=0.32]{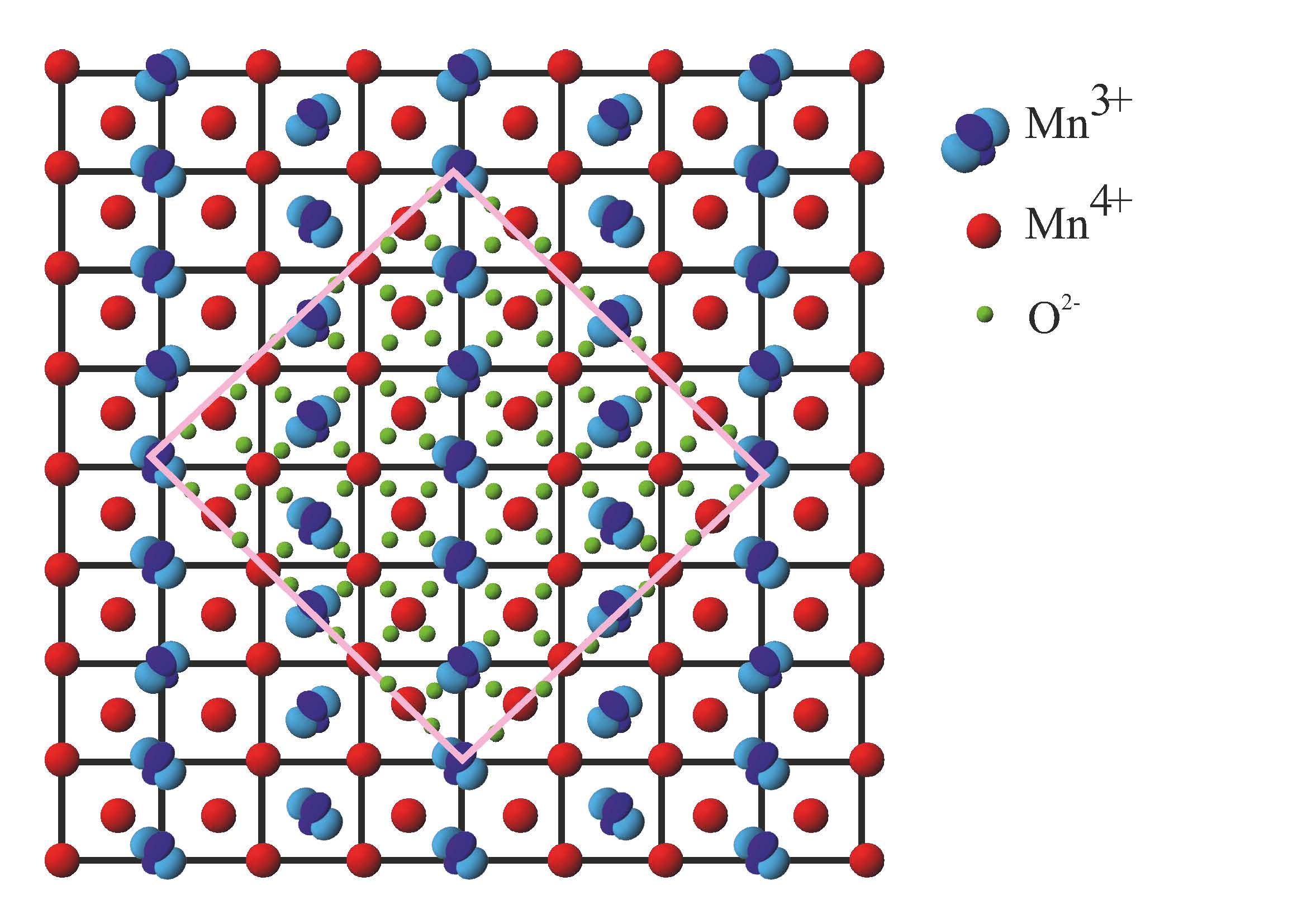}
\par\end{centering}

\caption{Possible OO ground state in the $ab$ plane for $\nicefrac{2}{3}$
doping in Tb$_{0.66}$Ca$_{0.33}$BaMn$_{2}$O$_{6}$. The unit cell
is indicated in the center of the double layer of Mn octahedra and
includes the oxygen ions. The Mn$^{3+}$ ions exhibit alternating
$x^{2}-z^{2}$ and $y^{2}-z^{2}$ orbitals.}

\end{figure}

In this paper we continue our investigations of the hole doped A-site
ordered manganites with neutron powder diffraction, SXRD and cluster
calculations in order to understand the alteration of the electronic
Mn states as function of doping and as function of temperature. The
crystal structures obtained with neutron powder diffraction reflect
the various orbital and charge ordered phases and show an increase
of the Mn-O-Mn bond angle as function of doping and temperature.

Theoretical simulations of the OO reflections $\left(\delta,\delta,0\right)$
recorded at the Mn $L_{2,3}$ edges \citep{Garcia-Fernandez-Sm-2008,Garcia-Fernandez2009}
for the various hole doping concentrations and as function of temperature
are presented. The cluster calculations consider the effect of the
bandwidth, hole doping and magnetic correlations and give insight
into the electronic ordering of the Mn-O hybridized states.

\begin{figure}[t]
\begin{centering}
\includegraphics[scale=0.4]{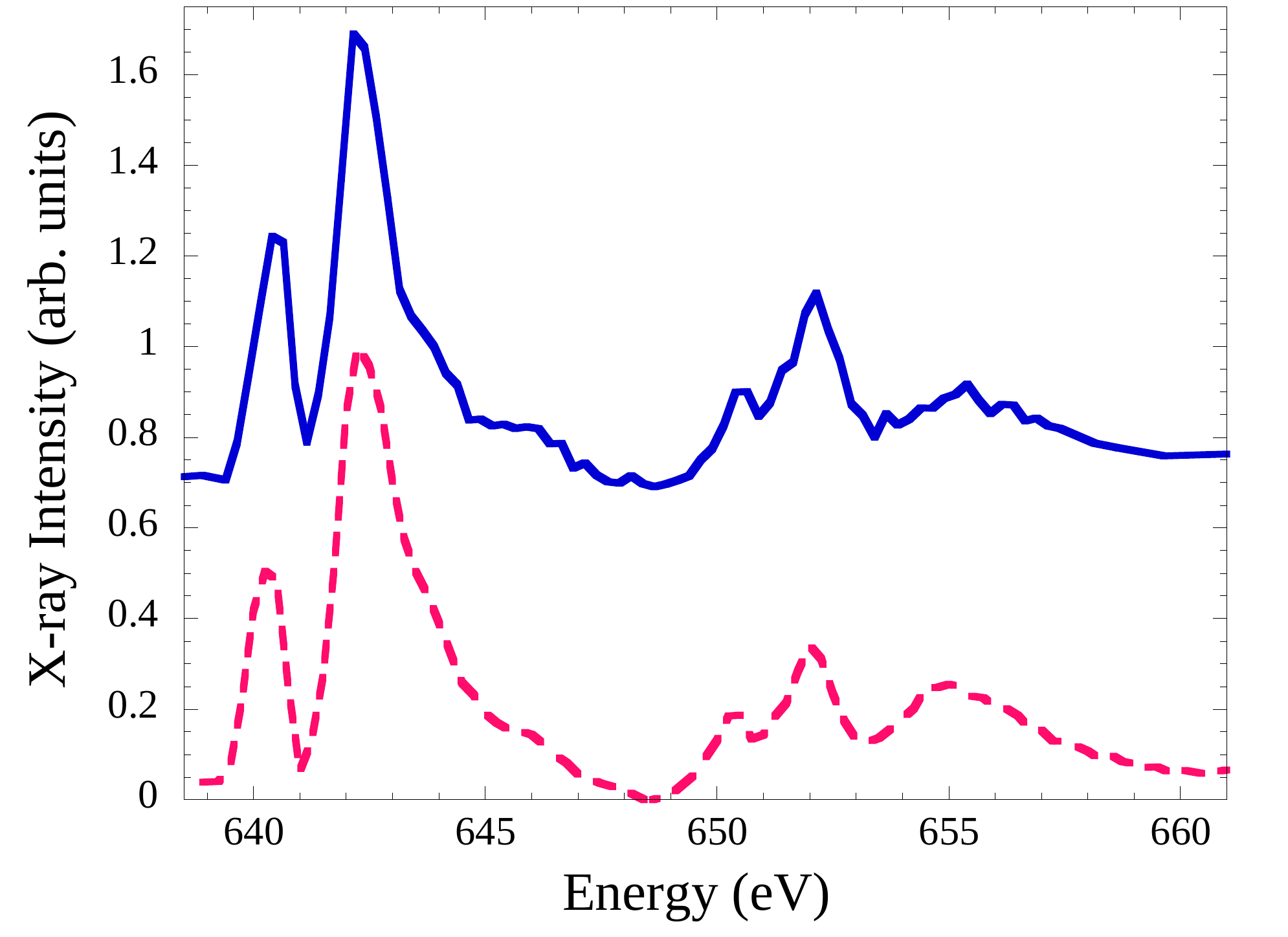}
\par\end{centering}

\caption{Energy dependence of the integrated intensity of the orbital reflection
($\nicefrac{1}{4}$, $\nicefrac{1}{4}$, 0) , measured at the Mn \textit{L}-edges
for polycrystalline Nd$_{0.4}$Tb$_{0.6}$BaMn$_{2}$O$_{6}$ (solid
line) at 26 K and SmBaMn$_{2}$O$_{6}$ \citep{Garcia-Fernandez-Sm-2008}
(striped line) at 120 K. Both sets of data were collected with $\pi$
incident radiation, and normalized ans shifted for clarity.}

\end{figure}

\begin{figure*}
\begin{centering}
\includegraphics[scale=0.6]{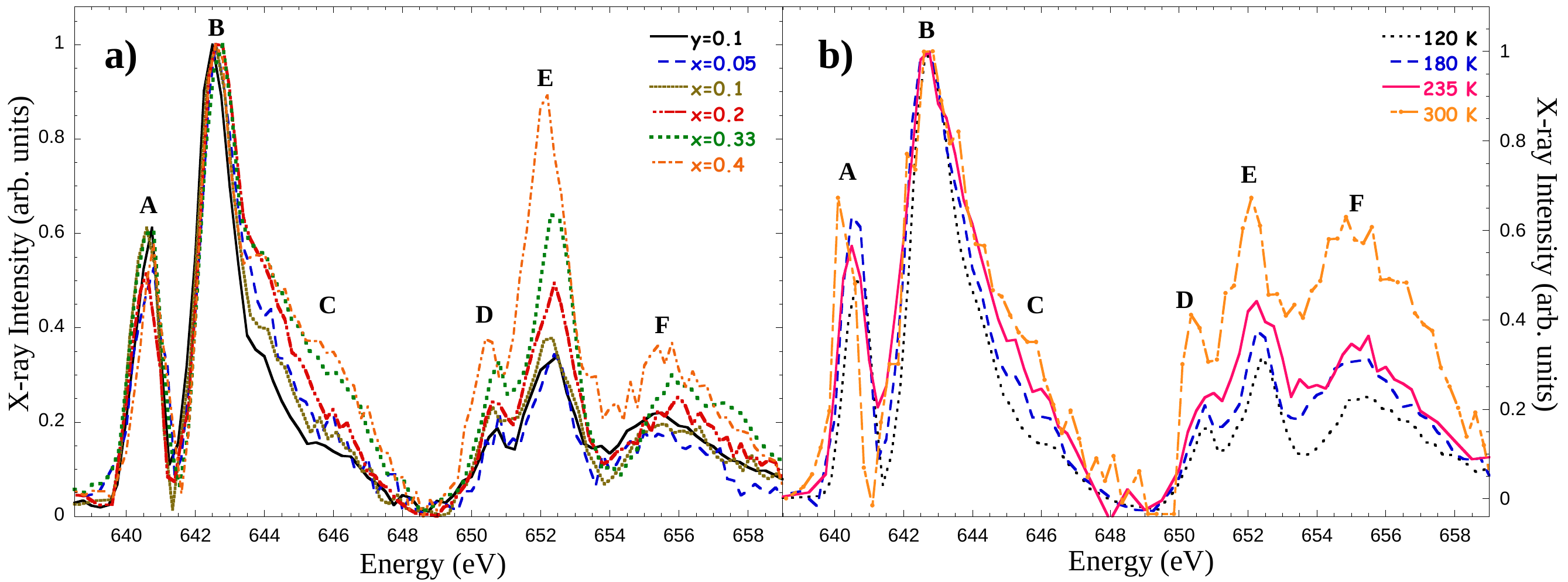}
\par\end{centering}

\caption{(color online) (a) Energy dependence of the integrated intensity of
the orbital reflection ($\delta_{x},$$\delta_{x},$0), measured at
the Mn \textit{L}-edges for systems above half doping Tb$_{1-x}$Ca$_{x}$BaMn$_{2}$O$_{6}$
and below half doping TbLa$_{y}$Ba$_{1-y}$Mn$_{2}$O$_{6}$ in polycrystalline
samples\citep{Garcia-Fernandez2009}. The experimental data have been
normalized to the intensity at the $L_{3}-$edge (643 eV) for comparison.
(b) Energy dependence of the integrated intensity of the ($\nicefrac{1}{4}$,
$\nicefrac{1}{4}$, 0) reflection at half doping of polycrystalline
SmBaMn$_{2}$O$_{6}$ with $\pi$ incident radiation, measured at
120 K, 180 K, 235 K and 300 K, renormalized to the intensity at the
$L_{3}$-edge (643 eV) \citep{Garcia-Fernandez-Sm-2008}.}

\end{figure*}

\begin{figure}
\begin{centering}
\includegraphics[scale=0.39]{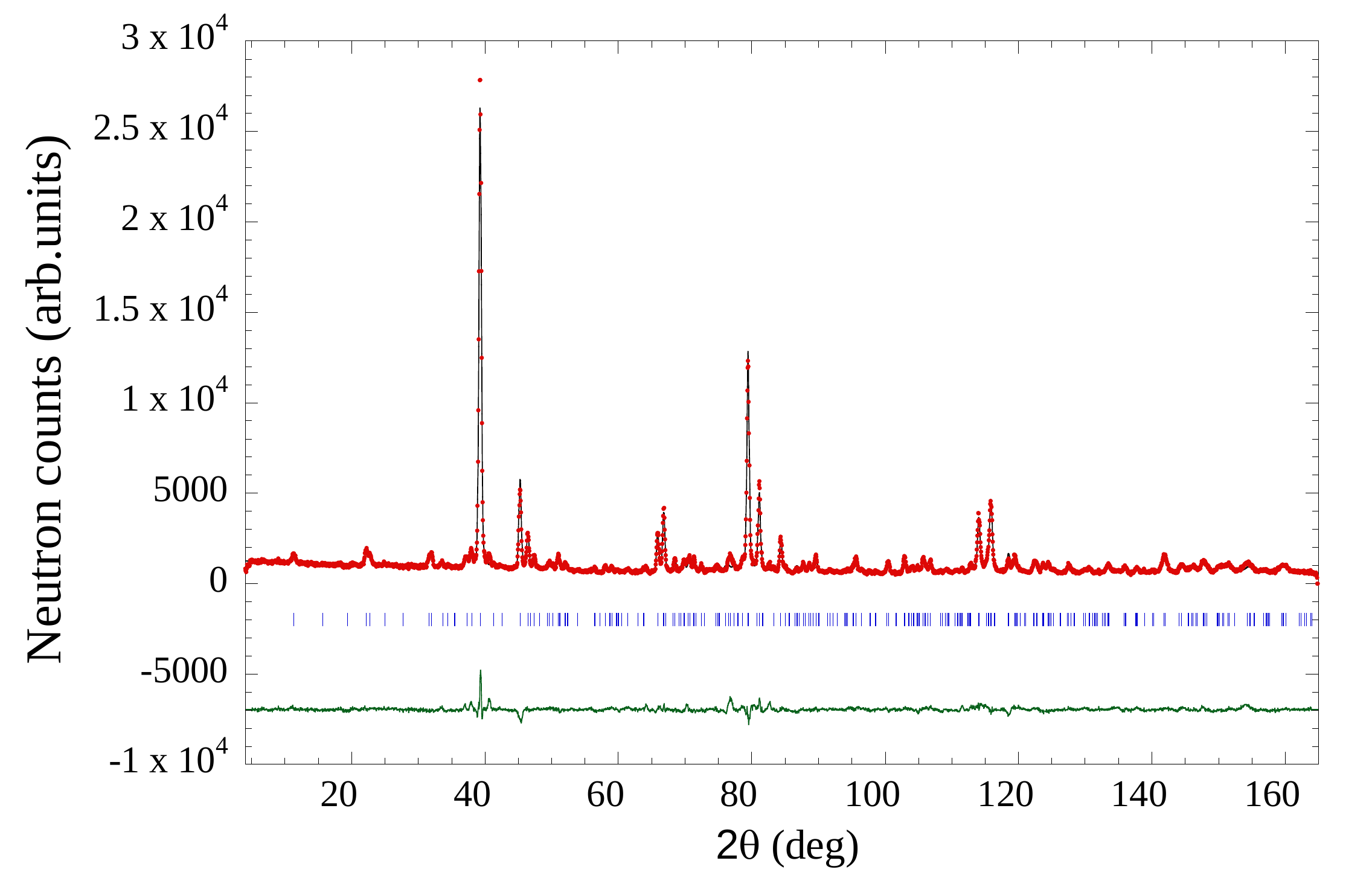}
\par\end{centering}

\caption{The Rietveld refinement pattern and difference plot of the neutron
diffraction data for the sample Tb$_{0.66}$Ca$_{0.33}$BaMn$_{2}$O$_{6}$
at T=300 K measured at HRPT with the wavelength $\lambda$= 1.49 $\textrm{\AA}$.
The rows of ticks show the Bragg peak positions. The structural parameters
are shown in Table 1.}

\end{figure}

\section{Experiments}

We have investigated polycrystalline samples of Tb$_{1-x}$Ca$_{x}$BaMn$_{2}$O$_{6}$
with doping contents $x$=0.05, 0.1, 0.2, 0.33 and 0.4, TbBa$_{1-y}$La$_{y}$Mn$_{2}$O$_{6}$
with $y$=0.1, and the half doped materials SmBaMn$_{2}$O$_{6}$
and Nd$_{0.4}$Tb$_{0.6}$BaMn$_{2}$O$_{6}$. The compounds were
synthesized following the same procedure as in ref \citep{Garcia-Fernandez-Sm-2008}.
A detailed structural investigation was performed on compositions
Tb$_{0.66}$Ca$_{0.33}$BaMn$_{2}$O$_{6}$ and Nd$_{0.4}$Tb$_{0.6}$BaMn$_{2}$O$_{6}$
using neutron powder diffraction. The measurements were carried out
at the high-resolution diffractometer for thermal neutrons, HRPT \citep{Fischer_2000}
at the SINQ neutron spallation source of the PSI, Switzerland. The
normal intensity mode of HRPT was used with the neutron wavelength
=1.49 $\textrm{\AA}$. All the temperature scans were carried out
on heating. The refinements of the crystal structure parameters were
performed using FULLPROF \citep{Rodriguez-Carvajal1993}.

Resonant soft x-ray diffraction experiments were performed at the
RESOXS \citep{Staub2008} endstation at the SIM beamline of the Swiss
Light Source of the Paul Scherrer Institut (PSI), Switzerland. Polycrystalline
pellets of the manganite with of 10 mm diameter were glued onto a
copper sample holder mounted on a He flow cryostat which achieves
temperatures between 10 K and 400 K. Experiments were performed using
linear horizontal or vertical polarized x-rays leading to $\pi$ or
$\sigma$ incident photon polarization in the horizontal scattering
geometry, respectively. Two-dimensional data sets were collected with
a commercial Roper Scientific charge-coupled-device (CCD) camera mounted
in vacuum. The sections of the resonant orbital powder diffraction
rings measured with the CCD camera are integrated along the vertical
direction and fitted with a pseudo-voigt function. This results in
the integrated intensity, the position in 2$\theta$ and the background,
which is mainly due to the fluorescence.

\section{Results}

The aim of this study is to compare the energy dependencies of the
resonant soft x-ray diffraction data from present A-site ordered manganites
(Figures 3 and 4) with theoretical simulations and structural investigations.
Since the neutron diffraction investigation of SmBaMn$_{2}$O$_{6}$
is hampered by the large neutron absorption coefficient of Sm, we
replace this cation by a mixture of Nd and Tb of the same average
radius. The crystal structure of resulting compound Nd$_{0.4}$Tb$_{0.6}$BaMn$_{2}$O$_{6}$
is determined by neutron powder diffraction without large absorption.
Although the tolerance factor in SmBaMn$_{2}$O$_{6}$ remains to
be slightly different than in Nd$_{0.4}$Tb$_{0.6}$BaMn$_{2}$O$_{6}$,
this produces only a small difference in the temperature at which
the MI transition takes place. This temperature corresponds to T$_{CO/OO}$$\approx$360
K for SmBaMn$_{2}$O$_{6}$ and T$_{CO/OO}$$\approx$400 K for Nd$_{0.4}$Tb$_{0.6}$BaMn$_{2}$O$_{6}$.
Figure 3 shows the energy dependence of the OO reflection $\left(\nicefrac{1}{4},\nicefrac{1}{4},0\right)$
measured in the vicinity of the Mn $L_{2,3}$ edges for both half
doped compounds. The data is normalized at E=643 eV to unity and shifted
vertically for comparison. The spectral shape of the orbital reflection
$\left(\nicefrac{1}{4},\nicefrac{1}{4},0\right)$ is identical within
the experimental uncertainty. This shows that the electronic ground
state in both compounds can be assumed to be the same, which makes
a direct comparison between structural data obtained by neutrons and
the spectroscopic x-ray data feasible.

The crystal structure of half doped $\mathit{R_{1-x}}T_{x}$MnO$_{3}$
manganites for which $R$ and $T$ ions are randomly distributed (solid
solution), such as La$_{0.5}$Ca$_{0.5}$MnO$_{3}$, can be described
in orthorhombic symmetry at high temperatures $\left(T\geq T_{CO/OO}\right)$.
The space group of this crystal structure is $Pnma$, which contains
only one symmetry equivalent Mn site and therefore does not describe
long-range CO at Mn sites \citep{Radaelli1997}. Below $T_{CO/OO}$,
the crystal structure is refined in a monoclinic $P2_{1}/m$ symmetry
$2\sqrt{2a_{p}}\,\mathrm{x}\,2a_{p}\,\mathrm{x}\,\sqrt{2a_{p}}$ superstructure,
with two inequivalent Mn sites in the unit cell, consistent with a
charge and orbital ordering \citep{Goff2004}. The structures of several
other half-doped manganites have subsequently been fitted using the
same model \citep{Blasco1997,Woodward1999,Richard1999}.

\begin{table*}[t]

\begin{centering}
\begin{tabular}{|c|c|c|c|c|c|}
\hline 
Atom  & Position  & x  & y  & z  & B\tabularnewline
\hline
\hline 
Tb  & 4h  & 0.243(1)  & 0  & 0.5  & 0.92(7)\tabularnewline
\hline 
Ca  & 4h  & 0.243(1)  & 0  & 0.5  & 0.92(7)\tabularnewline
\hline 
Ba  & 4g  & 0.246(1)  & 0  & 0  & 0.51(7)\tabularnewline
\hline 
Mn  & 8n  & 0  & 0.247(1)  & 0.2563(6)  & 0.49(5)\tabularnewline
\hline 
O1  & 4j  & 0  & 0.2069(7)  & 0.5  & 1.5(2)\tabularnewline
\hline 
O2  & 4i  & 0  & 0.263(1)  & 0  & 1.10(9)\tabularnewline
\hline 
O3  & 4l  & 0  & 0.5  & 0.2856(9)  & 1.3(1)\tabularnewline
\hline 
O4  & 4k  & 0  & 0  & 0.2446(7)  & 1.5(1)\tabularnewline
\hline 
O5  & 8m  & 0.25  & 0.25  & 0.2807(7)  & 1.73(8)\tabularnewline
\hline
\end{tabular}
\par\end{centering}

\caption{Structure parameters in Tb$_{0.66}$Ca$_{0.33}$BaMn$_{2}$O$_{6}$
{[}space group Cmmm (No.65){]}. The data are refined from the powder
neutron diffraction pattern measured at HRPT/SINQ with wavelength
$\lambda$=1.49$\textrm{\AA}$ (Fig. 5). The Bragg reliability factor
is R$_{Bragg}$=11.3\% and the conventional reliability factors are
R$_{wp}$=8.95, R$_{exp}$=4.70 and $\chi$$^{2}$=3.62.}

\end{table*}

\begin{figure*}
\begin{centering}
\includegraphics[scale=0.55]{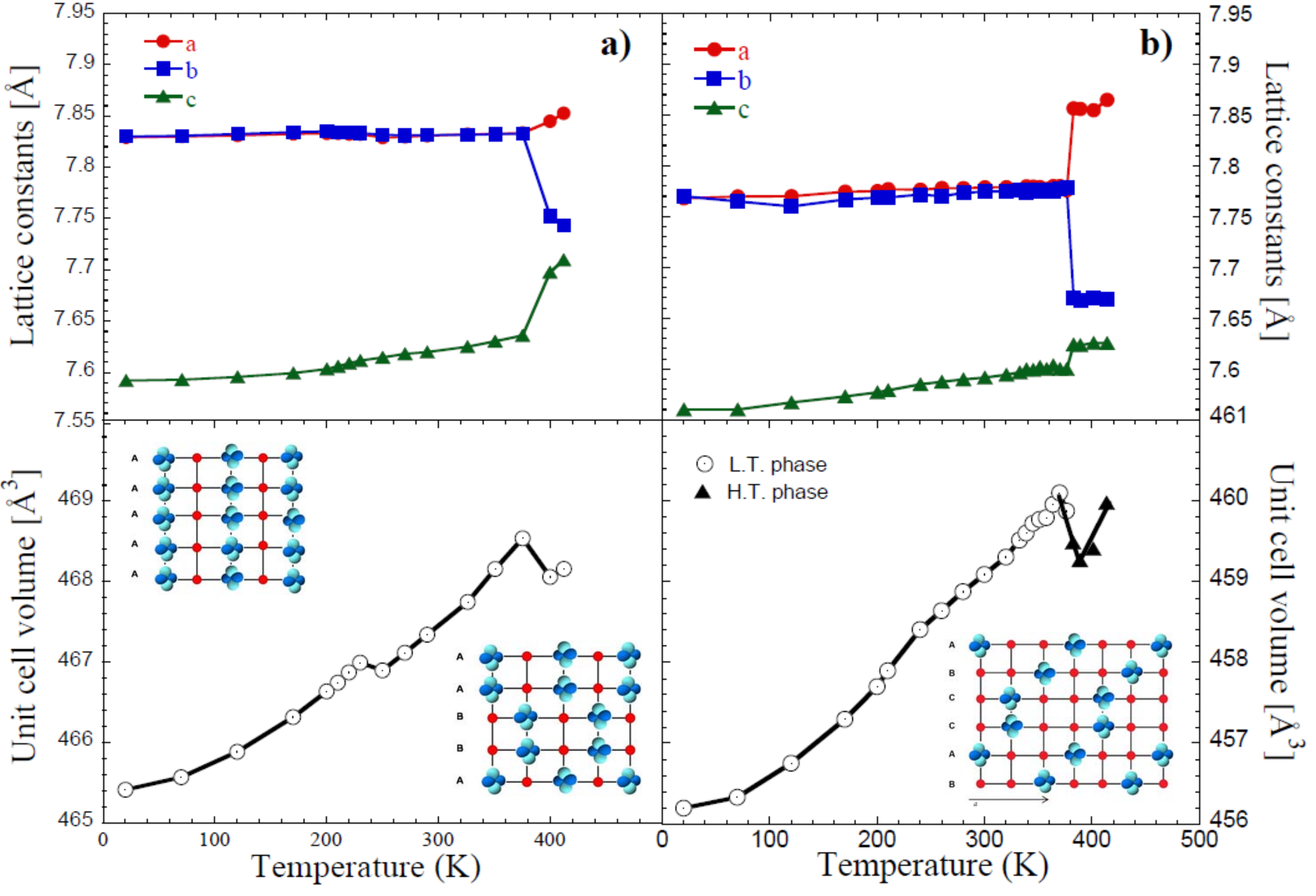}
\par\end{centering}

\caption{(a) (above) Unit cell parameters as a function of temperature for
Nd$_{0.4}$Tb$_{0.6}$BaMn$_{2}$O$_{6}$. (below) Variation of volume
with temperature for Nd$_{0.4}$Tb$_{0.6}$BaMn$_{2}$O$_{6}$. (b)
(above) Unit cell parameters as a function of temperature for Tb$_{0.66}$Ca$_{0.33}$BaMn$_{2}$O$_{6}$.
(below) Variation of volume with temperature for Tb$_{0.66}$Ca$_{0.33}$BaMn$_{2}$O$_{6}$
\citep{Garcia-Fernandez2009}. L.T. stands for low temperature phase
and H.T. for high temperature phase.}

\end{figure*}

\begin{figure*}
\begin{centering}
\includegraphics[scale=0.5]{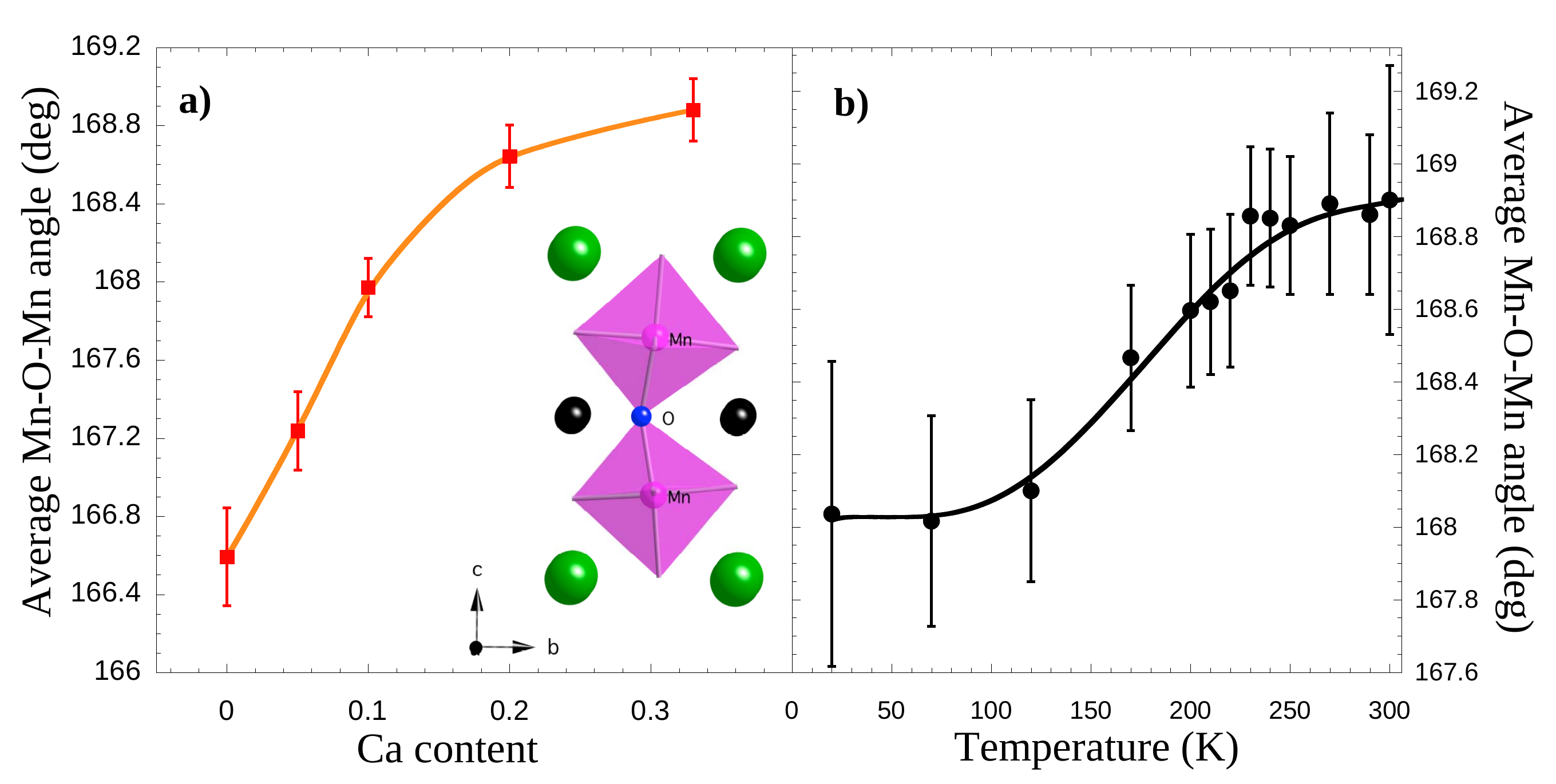}
\par\end{centering}

\caption{(a) Evolution of the average angle Mn-O-Mn as function of doping content
in Tb$_{1-x}$Ca$_{x}$BaMn$_{2}$O$_{6}$. The inset shows the unit
cell of the high-temperature structure of Tb$_{1-x}$Ca$_{x}$BaMn$_{2}$O$_{6}$
in which the Mn-O1-Mn angle is visualized as a tilt between the two
neighbouring oxygen octahedra. (b) Evolution of the average angle
between Mn-O-Mn as a function of temperature in Nd$_{0.4}$Tb$_{0.6}$BaMn$_{2}$O$_{6}$. }

\end{figure*}

As concerns the A-site ordered manganites like TbBaMn$_{2}$O$_{6}$,
their high temperature crystal structure is described by the orthorhombic
space group $Cmmm$. Similarly to solid solutions $R_{1-x}T_{x}$MnO$_{3}$,
all Mn sites are equivalent \citep{Williams2005}. At T$_{CO/OO}$=473
K, the system undergoes a crystal structure transition from orthorhombic
to monoclinic $P2_{1}/m$, associated with a diversification of Mn
valences or formation of Zener polarons \citep{Williams2005,Daoud-Aladine2008}.
However, the major monoclinic $P2_{1}/m$ superstructure reflections
were observed to be very weak in our measured powder patterns. Therefore
the neutron diffraction data were refined with orthorhombic (\textit{Cmmm})
crystal symmetry and $2a_{p}\mathrm{x}2a_{p}\mathrm{x}2a_{p}$ crystal
structure \citep{Rodriguez-Carvajal1993} for all temperatures and
doping contents. It was observed that when the stoichiometry of the
system changes, the diffraction pattern changes too. However, this
changes do not result in a detectable new crystal symmetry, but can
simply be described by the atomic motions and changes of the unit
cell parameters. The diffraction pattern at 300 K and the refinement
plot for the Tb$_{0.66}$Ca$_{0.33}$BaMn$_{2}$O$_{6}$ sample are
shown in Figure 5. The structure parameters obtained from this refinement
are shown in Table 1.

The MI transition is a first order transition leading to the coexistence
of two phases in the vicinity of the transition temperature. The temperature
dependence of the unit cell parameters and volume are shown in Figure
6 for the half doped compound Nd$_{0.4}$Tb$_{0.6}$BaMn$_{2}$O$_{6}$
and $\nicefrac{2}{3}$ doped compound Tb$_{0.66}$Ca$_{0.33}$BaMn$_{2}$O$_{6}$.
The lattice constants undergo dramatic changes. Two transitions can
be distinguished in the temperature evolution of the unit cell parameters
and the volume, one in the vicinity of $T{}_{CO/OO}\thicksim$410
K, (Figure 6) and the other in the vicinity of $T_{CO2}\thicksim$210
K .

When increasing the temperature above $T_{CO/OO}=410$ K (see figure
6a) in the half doped material, the unit cell parameters \textit{a}
and \textit{c} show sharp positive jumps at $T_{CO/OO}$ of 0.13\%
and 0.81\% respectively, while \textit{b} drops by -1.05\%. Those
anisotropic changes of the crystal unit cell result in an abrupt drop
of the unit cell volume at $T_{CO/OO}$ by -0.11\%. A similar volume
collapse was also observed at the MI transition in A-site ordered
HoBaCo$_{2}$O$_{5.5}$ \citep{key-69}. In this cobaltite, the volume
collapse of the unit cell has been associated to the occurrence of
orbital order of the Co$^{3+}$ ions. In the vicinity of $T_{CO2}=210$
K, where a restacking of the OO has been observed \citep{Garcia-Fernandez-Sm-2008,Akahoshi2006},
a small drop in the unit cell volume is observed.

\begin{figure}
\begin{centering}
\includegraphics[scale=0.31]{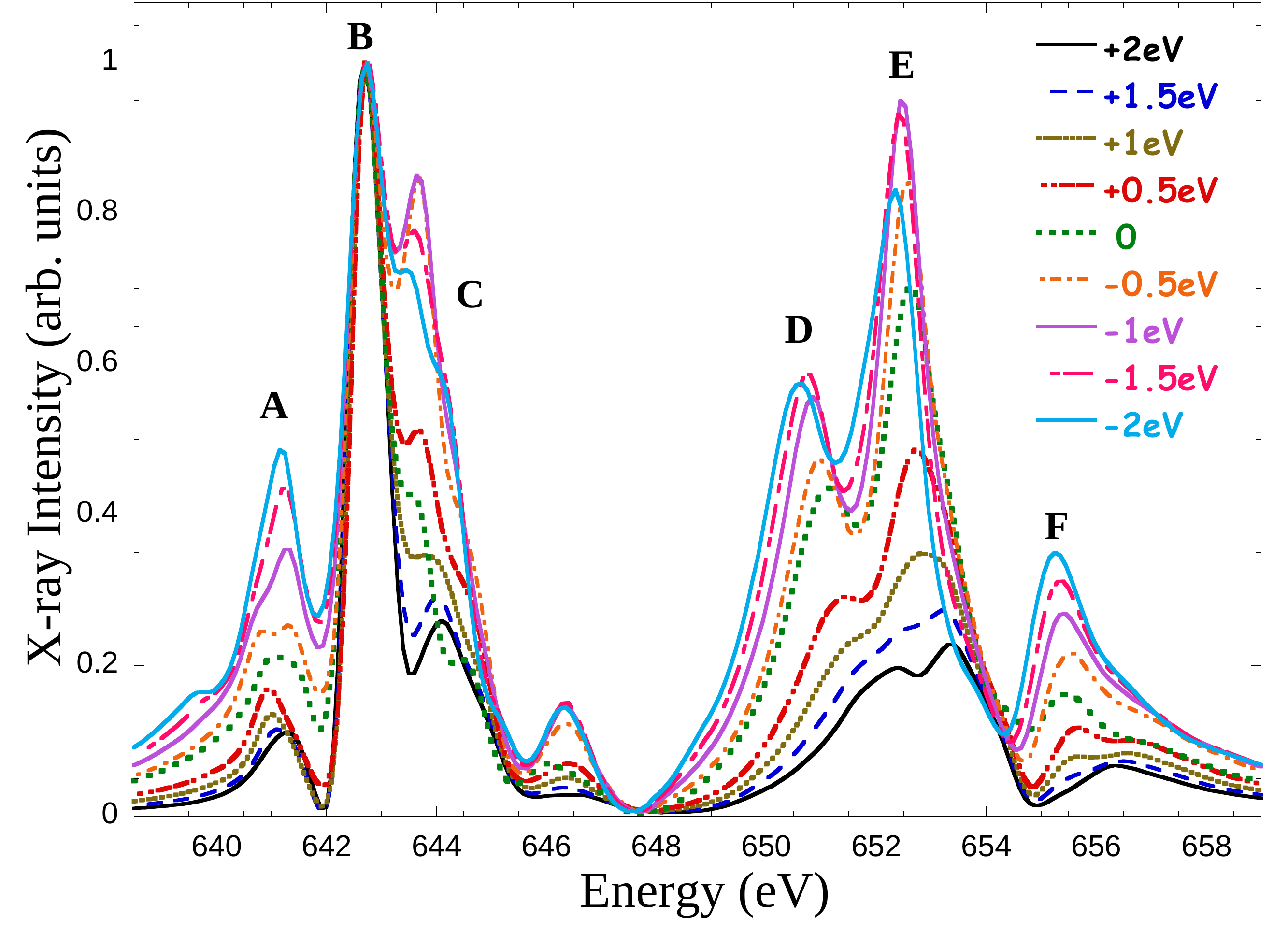}
\par\end{centering}

\caption{Change in spectral shape of the $\left(\delta_{x},\,\delta_{x},\,0\right)$
reflection as a function of $\varepsilon_{p}$ which results in a
variation of the hole density at the oxygen sites.}

\end{figure}

Similar behaviour is observed close to the MI transition in $\nicefrac{2}{3}$
doped Tb$_{0.66}$Ca$_{0.33}$BaMn$_{2}$O$_{6}$ (see figure 6b).
The unit cell parameters \textit{a} and \textit{c} show a sharp increase
at $T_{CO/OO}$ of 0.97\% and 0.30\% respectively, while \textit{b}
decreases by -1.41\%. The anisotropic changes of the crystal unit
cell result in an abrupt drop of the unit cell volume at $T_{CO/OO}$
of -0.16\%. Surprisingly, the changes of the lattice constants \textit{a}
and \textit{b} are larger than in the half doped system. In contrast,
in the vicinity of $T_{CO2}$ = 210 K, a decrease in unit cell volume
as in the case of Nd$_{0.4}$Tb$_{0.6}$BaMn$_{2}$O$_{6}$ is absent,
yet a change in slope is noted. This agrees with the fact that OO
restacking does not occur for the hole doped compounds. We note that
additional reflections are observed for $T<200K$, indicative for
antiferromagnetic ordering with $T_{N}<T_{CO2}$.

In summary, a similar contraction of the crystal structure is seen
for all hole doped systems near the MI transition at $T_{CO/OO}$,
whereas the changes in crystal structure near $T_{CO2}$ are different.

\section{Discussion}

Previously the energy dependence of the orbital reflection $\left(\delta_{x},\delta_{x},0\right)$,
where $\left[\delta_{x}=\delta\left(x\right)\right]$, for $x>0$
and $\left(\nicefrac{1}{4},\nicefrac{1}{4},0\right)$ for $y=0.1$,
in the vicinity of the Mn $L_{2,3}$ edges was recorded for different
doping contents close to half doping (ref. \citep{Garcia-Fernandez2009}
reproduced in Figure 3.a). The spectra are normalized to their maxima
at the Mn $L_{3}$ edge. It shows that there is a distinct trend in
the shape of the spectra with doping, i.e. some features of the energy
become broader or more intense compared with others. Theses differences
indicate a variation of the electronic Mn states. In a simple view
such a change is not expected. When the doping is increased beyond
half doping, the holes are disposed as far as possible from each other,
giving rise to the linear behavior of \textit{q} versus doping \citep{Garcia-Fernandez2009}.
In this case, the local structure around the Mn$^{3+}$ ions (considering
the nearest neighbors only) is not directly affected by doping.

In Figure 3.b the energy dependence of the OO reflection $\left(\nicefrac{1}{4},\nicefrac{1}{4},0\right)$
measured in the vicinity of the Mn $L_{2,3}$ edges for the half doped
material SmBaMn$_{2}$O$_{6}$ is shown for temperatures between 120
K and 300 K normalized at E=643 eV to unity. In a previous study of
this half doped compound it was shown that the energy dependences
do not change dramatically with temperature \citep{Garcia-Fernandez-Sm-2008}
in contrast to what was observed in La$_{0.5}$Sr$_{1.5}$MnO$_{4}$
\citep{Staub2006}. In the latter, the different features of the energy
scan were shown to follow different trends as function of temperature,
which was interpreted in terms of different order parameters of orbital
order and Jahn-Teller distortion. Features labeled C and F correspond
to Jahn Teller distortion, features A and B mostly to the OO. Although
this clear different behavior is absent for SmBaMn$_{2}$O$_{6}$,
small but distinct dissimilarities of different type can also be seen
between energy dependences measured at different temperatures. Features
D, E and F become more intense upon heating, while feature C broadens
in comparison to features A and B. The evolution of the spectra as
a function of temperature follows a similar trend as the ones observed
as a function of doping.

The $L_{2,3}$ edges are sensitive to the local environment of the
scattering atom. Doping and temperature modify the local environment
structurally and/or electronically.

Concerning structural changes, in the \textit{Cmmm} space group and
with the 2$a_{p}$x2$a_{p}$x2$a_{p}$ cell we have five different
oxygen sites leading to several Mn-O-Mn angles, and different oxygen
positions (see Table I) . To simplify the comparison we merge those
angles obtained from the structural refinements to obtain an average
Mn-O-Mn bond angle (Fig. 7a). As can be seen in Figure 7a, the average
Mn-O-Mn angle increases for increasing doping. The work of J.L. García-Muñoz
et al. \citep{Garcia-Munoz1996} indicates that this corresponds to
an increase in bandwidth. We plot in Figure 7b the analogous result
obtained as a function of increasing temperature, showing a similar
trend.

The change of the average angle as a function of doping is about twice
as large compared to the one upon temperature. The error bars versus
temperature obtained from statistical error propagation of the individual
errors are very large compared to the scattering of the data points.
This indicates that the obtained individual bond angles are significantly
correlated. Therefore the accuracy of the average angle is much better
than the plotted error bars obtained from statistical error propagation.

Another interesting point is the relation between the orbital restacking
transition at half-doping and the anomalous orbital melting for the
$\nicefrac{2}{3}$ doping. Both transitions, which have been previously
observed in soft x-ray resonant scattering measurements, can also
be detected in the temperature dependence of the unit cell volume
of both compounds (Figure 6). For the half-doping case, at low temperatures
the orbital planes (Mn$^{3+}$) are aligned along the \textit{c}-axis,
whereas above T$_{CO2}$, the stacking of the orbital planes is pair
wise shifted in the plane. This orbital stacking transition is sketched
in the inset of Figure 6a. For the $\nicefrac{2}{3}$ doping, at low
temperatures we could assume again a ferro-type staking along the
\textit{c}-axis, but above T$_{CO2}$, differently to the previous
case, there's much more freedom to stack the individual planes along
the \textit{c}-axis. Shifting a plane perpendicularly to \textit{c}
by one perovskite unit cell, in one direction or its opposite, does
not lead to the same result (see inset Figure 6b). The extended possibilities
of having different stacking along the \textit{c}-axis, lead to many
states which differ very little in energy. This allows having a thermally
excited switching between the different stacking representing a dynamical
sliding of the planes and that is equivalent to a correlated motion
of electrons. This decoupling of ordered planes leads then to the
slight volume increase above the onset of the dynamics that is visible
in Figure 6b. Moreover, such dynamical sliding will also be expected
to affect the correlation in the ab plane. The change of ordering
period in this dynamical regime might then reflect a slightly smaller
doping than the one expected from the exact Ca content. The OO unit
cell is free to relax in the plane due the diminishing coupling along
the \textit{c} axis for increasing temperatures. Such sliding planes
might possibly be related to the recently proposed sliding character
of the charge density\citep{Cox2008}.

Calculations to describe the spectral shape of the orbital reflection
have been done using the model of reference \citep{Mirone2006}. This
model describes, in second quantization, a small cluster consisting
of a central Mn$^{3+}$ site and the first neighbouring shells of
O and Mn$^{4+}$ sites. The interaction term $T{}_{1}$ of equation
2 of reference \citep{Mirone2006} has been modified in order to obtain
an orbital ordering of the $e_{g}$ electron of $x^{2}-y^{2}$ kind
in the ground state which corresponds to the presumed ordering for
our physical system \citep{Garcia-Fernandez-Sm-2008}. Here we like
to note that the calculation of the structure factor does not depend
directly on the ordering wave vector change for increased doping,
as the scattering factors are calculated in the cluster code. It merely
is reflected by the phase difference of $\pi$ of the two orbital
oriented sites $\left(\nicefrac{x^{2}-z^{2}}{y^{2}-z^{2}}\right)$
in the resonant x-ray structure factor. The $T{}_{1}$ term hybridizes
the Mn 3\textit{d} orbitals with the 2\textit{p} orbitals of the neighbouring
oxygen ions and $T_{1}$ is given by:

\begin{equation}
T_{1}=\sqrt{2}\, t\sum_{\sigma}\left(g\, o_{x,\sigma}^{\dagger}d_{x2,\sigma}+o_{z,\sigma}^{\dagger}d_{z2,\sigma}+go_{y,\sigma}^{\dagger}d_{y2,\sigma}\right)+\mathrm{c.c.},\end{equation}

where the symbols $o$ and $d$ are second quantization operators
for the oxygen and central Mn orbitals respectively. The $y$ axis
is in the crystal $c$ direction. The value $g=0.7$ from reference
\citep{Mirone2006} is used. In the original model the $g$ factor
multiplies only the $o_{x,\sigma}^{\dagger}d_{x2,\sigma}$ term, thus
favoring a $3x^{2}-r^{2}$ orbital ordering. Here, the $g$ factor
multiplies also the $o_{y,\sigma}^{\dagger}d_{y2,\sigma}$ hopping
operator, favoring $x^{2}-y^{2}$ type OO.

The effect of change in the bandwidth on the OO spectra can be simulated
by changing the $t$ prefactor in the $T_{1}$ term and/or in the
$T_{2}$ one (equation 3 of reference \citep{Mirone2006}) which hybridizes
the 2\textit{p} oxygen orbitals with the Mn$^{4+}$ ones. By varying
$t$ proportionally to the cosine of the Mn-O-Mn angle from Figure
7 we did not observe noticeable effects. An angle variations of the
order of 1\% has a too small effect on the bandwidth. This indicates
that the change in bandwidth does not directly reflect the changes
of the spectral shape of the orbital reflection. Correspondingly,
it might be a secondary effect caused by the doping and not relevant
in the understanding of the electronic changes versus doping.

The direct effect of doping on the electronic structure has been simulated
in the hypothesis of hole density concentrated on oxygen sites. To
do so we have varied the $\varepsilon_{p}$ parameter in equation
4 of reference \citep{Mirone2006} to take into account the electrostatic
interaction of holes on the oxygen orbitals. $\varepsilon_{p}$ represents
the bare energy of the oxygen $p$ orbitals, and is deduced from the
charge transfer energy $\Delta$, which is defined as the energy to
transfer an electron from the oxygen onto a bare Mn atom (in the absence
of hybridization). Lowering $\varepsilon_{p}$ reflects therefore
a transfer (or condensation) of holes at the oxygen. The oxygen U$_{pp}$
Hubbard term being estimated to 5 eV (reference \citep{Mirone2006}),
an increase in x of 0.4 corresponds to a decrease in $\varepsilon_{p}$
of 1 eV in the case of holes going to in-plane oxygen sites.

\begin{figure}
\begin{centering}
\includegraphics[scale=0.45]{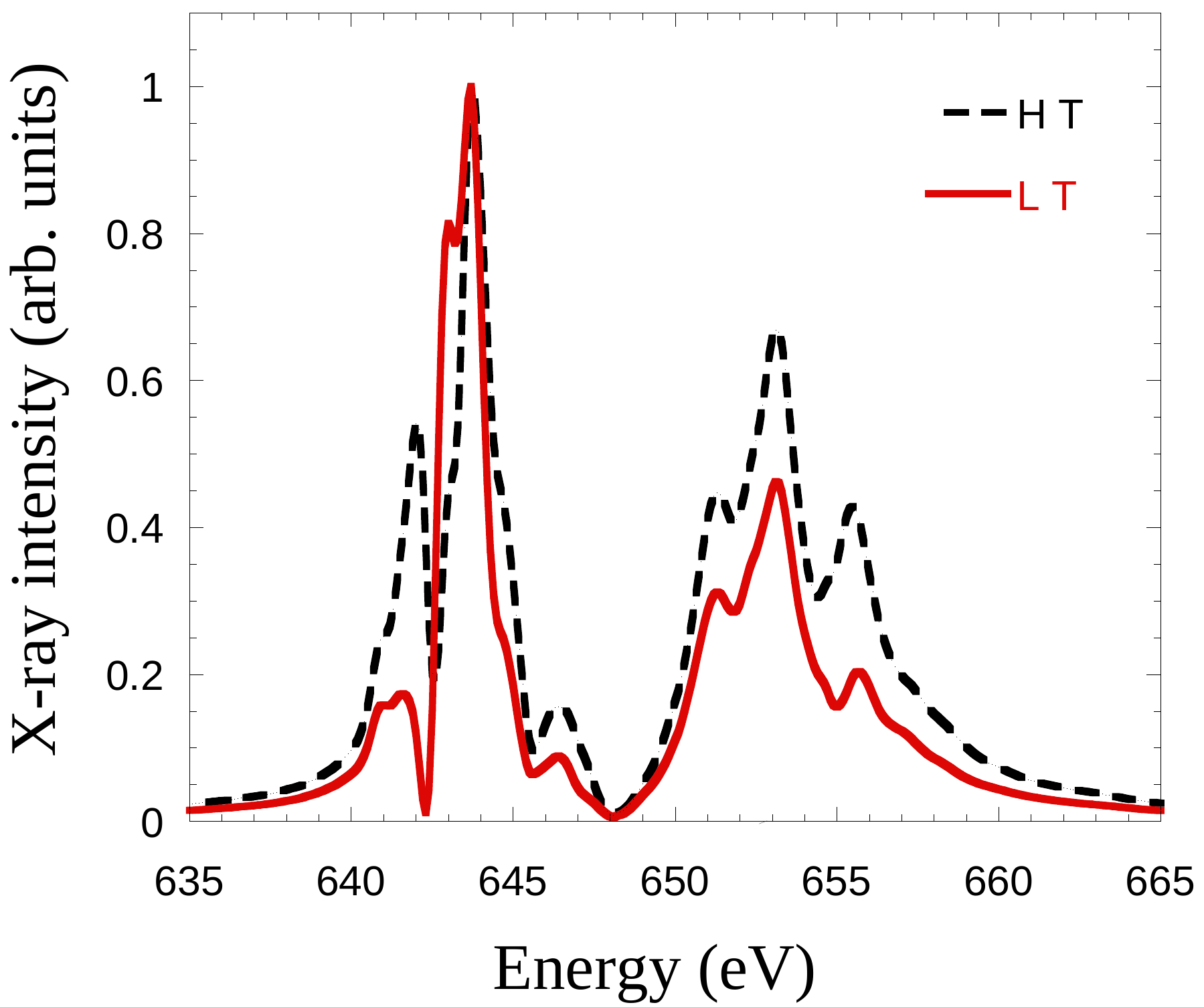}
\par\end{centering}

\caption{Theoretical simulation of the effect of changing the magnetic correlations
on the energy dependence of the orbital reflection $\left(\delta_{x},\,\delta_{x},\,0\right)$
in the vicinity of Mn \textit{L}$_{2,3}$-edge.}

\end{figure}

Figure 8 shows the change in spectral shape of the \textcolor{black}{$\left(\delta_{x},\delta_{x},0\right)$}
reflection as a function of $\varepsilon_{p}$. A decrease of the
energy of the oxygen 2\textit{p} orbitals leads to an increase of
the $L_{2}$ edge and of feature C. The agreement between these calculations
in Figure 8 and the experimental result obtained as function of hole
doping (Figure 3a) is prominent.

We have also tested the hypothesis of hole density increase at the
Mn ion site by changing the bare energy of the Mn $d$ orbitals $\varepsilon_{d}$.
These simulations of increased holes at the Mn sites show an opposite
trend in the energy dependencies compared to those calculated for
increased holes at the oxygen sites. We conclude therefore that the
holes are introduced in the oxygen 2\textit{p} shell rather than in
the Mn 3\textit{d} shell. The fact that the hole doping leads to a
linear dispersion of the orbital reflection with a $\nicefrac{2}{3}$
doping being represented by a tripling of the unit cell, indicates
that the holes are not centered at the oxygen nor at the bond, but
are centered around the Mn ions. In other words, it suggests that
each neighboring oxygen gets the same fraction of the hole so that
the center of mass stays at the Mn site, but is not directly affecting
the 3\textit{d} electron count.

This view is consistent with very recent investigation of the Mn $K_{\beta}$
fluorescence spectroscopy \citep{Herrero-Martin2010} and theoretical
calculations based on LDA+U of the electronic structure of manganites
\citep{Volja2010}. They find that the holes are introduced at the
oxygen sites, forming an $e_{g}$ symmetry orbital, leaking charge
to the former Mn$^{4+}$ sites. This results in a reduced charge transfer
between the former Mn$^{3+}$ and Mn$^{4+}$ sites, representing a
strong covalent character of the Mn-O sublattice.

The temperature dependence of the energy dependence of the OO reflection
at half doping cannot be explained by an increase in hole density
at the oxygen site. Such increase of holes is expected to increase
the intensity at the oxygen \textit{\textcolor{black}{K}} edge, as
there are more empty states accessible as intermediate states in the
resonant process. In addition the intensity at the Mn $L_{3}$ main
edge (643 eV) is expected to decrease, as the total charge is temperature
independent. This is in contrast with the observed temperature dependence
of the OO intensity at the oxygen \textit{K} edge and Mn $L_{3}$
edge respectively, which are equal \citep{Garcia-Fernandez2009}.
Moreover, the observed increase in Mn-O-Mn angle, which remains small,
does not support the observed increase in relative intensity at the
$L_{2}$ edge. This suggest that the temperature effect on the spectral
shape of the orbital reflection is probably more complex. It has been
shown previously that the magnetic moment direction as well as spin
correlations changes the spectral shape of the orbital reflection
\citep{Mirone2006,Stojic2005}.

To investigate the influence of spin correlations on the spectral
shape at half doping we consider the lowest calculated energy eigenstates,
in particular the first five eigenstates. They are contained within
an energy range of the order of 10 meVs and correspond to different
orientations of the central Mn spin \textit{S}. They originate from
the polarising term $H_{2}$ \citep{Mirone2006} which acts on the
$e_{g}$ orbitals of the cluster Mn$^{4+}$ external ions. This polarising
term accounts, for the on-site mean-field exchange with the $t_{2g}$
orbitals. These outer Mn $t_{2g}$ orbitals, are absent from the model,
and are represented by this mean field term. The inclusion of the
external $t_{2g}$ states in the model would be computationally very
expensive.

Therefore we have performed two calculations. The first calculation,
named LT, has been performed as described above inducing magnetic
correlations $H_{2}$ as in reference \citep{Mirone2006} at low temperature.
The scattering factors from the first five eigenstates are averaged
with weights decreasing versus eigenenergy.

For the second calculation, named HT, we reduce the slope of the weights
dependence versus eigen-energy and at the same time we reduce the
polarising term by dividing \textit{h} by two in $H_{2}$ \citep{Mirone2006}.
The parameter $h$ is a direct measure of the spin correlations and
reducing it, mimics the effect of temperature induced spin disorder
and the consequent reduction of the Mn-Mn magnetic correlation. The
factors that we use describe qualitatively a temperature increase
from 120 K at LT to 300 K at HT. The result of these calculations
is shown in Figure 9. The spectral shape changes mainly by an increase
of intensity at the $L_{2}$ edge compared to the $L_{3}$ edge. Our
simulation reproduces the main trend in spectral shape giving further
support of the importance of magnetic correlations in the description
of the electronic properties of these manganites systems.

\section{Conclusions}

The hole doping and temperature dependence of A-site ordered manganites
has been studied with neutron powder diffraction. In addition, we
compare the resonant soft x-ray powder diffraction spectra to cluster
calculations. The orbital order of the Mn$^{3+}$ 3\textit{d }states
exhibits electronic changes as function of hole doping and as function
of temperature. We have explored the origin of these electronic changes.
The spectral changes observed in the orbital reflection with resonant
diffraction at the Mn $L_{2,3}$ edges resemble similarity as function
of hole doping and temperature. Our theoretical modeling demonstrates
that the relative intensity at the $L_{2}$ edge increases, both for
increased hole density at the oxygen and for induced spin disorder
as function of temperature. In case of hole doping, the resonant spectra
are well described with additional holes residing in the oxygen 2\textit{p}
shell yet centered around the Mn$^{3+}$ ions. In contrast, the electronic
changes observed as function of temperature are more likely caused
by spin correlations.

The structural study found a volume increase at the anomalous orbital
melting transition temperature for the $\nicefrac{2}{3}$ doping system,
interpreted in terms of a decoupling and an onset of a dynamical sliding
of two dimensional OO planes. However, including these changes of
the bandwidth in the model calculations do not lead to observable
changes of the calculated spectra, and therefore indicate that the
bandwidth change as a function of doping and temperature is a secondary
effect.

\section*{Acknowledgements}

The experiments were performed and at the X11MA beamline of the SLS
at the SINQ of the Paul Scherrer Institut, Villigen, Switzerland and
we want to thank their staff for their excellent support. We acknowledge
the financial support of the following institutions, the Swiss National
Science Foundation, the NCCR MaNEP project, the {}``Access to Major
Research Facilities Programme\textquotedblright{} which is a component
of the {}``International Science Linkages Programme\textquotedblright{}
established under the Australian Government's innovation statement,
{}``Backing Australia's Ability\textquotedblright{} NSFCH and the
Gobierno del Principado de Asturias for the financial support of a
Postdoctoral grant from Plan de Ciencia, Tecnologia e Innovacion (PCTI)
de Asturias 2006\textendash{}2009.

\bibliographystyle{apsrev4-1}
\bibliography{garcia}

%merlin.mbs 2010-03-15 4.21a (PWD, AO, DPC)
\begin{thebibliography}{30}%
\makeatletter
\providecommand \@ifxundefined [1]{%
 \@ifx{#1\undefined}
}%
\providecommand \@ifnum [1]{%
 \ifnum #1\expandafter \@firstoftwo
 \else \expandafter \@secondoftwo
 \fi
}%
\providecommand \@ifx [1]{%
 \ifx #1\expandafter \@firstoftwo
 \else \expandafter \@secondoftwo
 \fi
}%
\providecommand \natexlab [1]{#1}%
\providecommand \enquote  [1]{``#1''}%
\providecommand \bibnamefont  [1]{#1}%
\providecommand \bibfnamefont [1]{#1}%
\providecommand \citenamefont [1]{#1}%
\providecommand \href@noop [0]{\@secondoftwo}%
\providecommand \href [0]{\begingroup \@sanitize@url \@href}%
\providecommand \@href[1]{\@@startlink{#1}\@@href}%
\providecommand \@@href[1]{\endgroup#1\@@endlink}%
\providecommand \@sanitize@url [0]{\catcode `\\12\catcode `\$12\catcode
  `\&12\catcode `\#12\catcode `\^12\catcode `\_12\catcode `\%12\relax}%
\providecommand \@@startlink[1]{}%
\providecommand \@@endlink[0]{}%
\providecommand \url  [0]{\begingroup\@sanitize@url \@url }%
\providecommand \@url [1]{\endgroup\@href {#1}{\urlprefix }}%
\providecommand \urlprefix  [0]{URL }%
\providecommand \Eprint [0]{\href }%
\@ifxundefined \urlstyle {%
  \providecommand \doi  [0]{\begingroup \@sanitize@url \@doi}%
  \providecommand \@doi [1]{\endgroup \@@startlink {\doibase
  #1}doi:\discretionary {}{}{}#1\@@endlink }%
}{%
  \providecommand \doi  [0]{doi:\discretionary{}{}{}\begingroup
  \urlstyle{rm}\Url }%
}%
\providecommand \doibase [0]{http://dx.doi.org/}%
\providecommand \Doi [0]{\begingroup \@sanitize@url \@Doi }%
\providecommand \@Doi  [1]{\endgroup\@@startlink{\doibase#1}\@@Doi}%
\providecommand \@@Doi [1]{#1\@@endlink}%
\providecommand \selectlanguage [0]{\@gobble}%
\providecommand \bibinfo  [0]{\@secondoftwo}%
\providecommand \bibfield  [0]{\@secondoftwo}%
\providecommand \translation [1]{[#1]}%
\providecommand \BibitemOpen [0]{}%
\providecommand \bibitemStop [0]{}%
\providecommand \bibitemNoStop [0]{.\EOS\space}%
\providecommand \EOS [0]{\spacefactor3000\relax}%
\providecommand \BibitemShut  [1]{\csname bibitem#1\endcsname}%
%</preamble>
\bibitem [{\citenamefont {Efremov}\ \emph {et~al.}(2004)\citenamefont
  {Efremov}, \citenamefont {Van~den Brink},\ and\ \citenamefont
  {Khomskii}}]{Efremov2004}%
  \BibitemOpen
  \bibfield  {author} {\bibinfo {author} {\bibfnamefont {D.~V.}\ \bibnamefont
  {Efremov}}, \bibinfo {author} {\bibfnamefont {J.}~\bibnamefont {Van~den
  Brink}}, \ and\ \bibinfo {author} {\bibfnamefont {D.~I.}\ \bibnamefont
  {Khomskii}},\ }\href@noop {} {\bibfield  {journal} {\bibinfo  {journal}
  {Nature Materials},\ }\textbf {\bibinfo {volume} {3}},\ \bibinfo {pages}
  {853} (\bibinfo {year} {2004})}\BibitemShut {NoStop}%
\bibitem [{\citenamefont {Daoud-Aladine}\ \emph {et~al.}(2008)\citenamefont
  {Daoud-Aladine}, \citenamefont {Perca}, \citenamefont {Pinsard-Gaudart},\
  and\ \citenamefont {Rodriguez-Carvajal}}]{Daoud-Aladine2008}%
  \BibitemOpen
  \bibfield  {author} {\bibinfo {author} {\bibfnamefont {A.}~\bibnamefont
  {Daoud-Aladine}}, \bibinfo {author} {\bibfnamefont {C.}~\bibnamefont
  {Perca}}, \bibinfo {author} {\bibfnamefont {L.}~\bibnamefont
  {Pinsard-Gaudart}}, \ and\ \bibinfo {author} {\bibfnamefont {J.}~\bibnamefont
  {Rodriguez-Carvajal}},\ }\href@noop {} {\bibfield  {journal} {\bibinfo
  {journal} {Phys. Rev. Lett.},\ }\textbf {\bibinfo {volume} {101}},\ \bibinfo
  {pages} {166404} (\bibinfo {year} {2008})}\BibitemShut {NoStop}%
\bibitem [{\citenamefont {Garcia-Fernandez}\ \emph {et~al.}(2009)\citenamefont
  {Garcia-Fernandez}, \citenamefont {Staub}, \citenamefont {Bodenthin},
  \citenamefont {Scagnoli}, \citenamefont {Pomjakushin}, \citenamefont
  {Lovesey}, \citenamefont {Mirone}, \citenamefont {Herrero-Martin},
  \citenamefont {Piamonteze},\ and\ \citenamefont
  {Pomjakushina}}]{Garcia-Fernandez2009}%
  \BibitemOpen
  \bibfield  {author} {\bibinfo {author} {\bibfnamefont {M.}~\bibnamefont
  {Garcia-Fernandez}}, \bibinfo {author} {\bibfnamefont {U.}~\bibnamefont
  {Staub}}, \bibinfo {author} {\bibfnamefont {Y.}~\bibnamefont {Bodenthin}},
  \bibinfo {author} {\bibfnamefont {V.}~\bibnamefont {Scagnoli}}, \bibinfo
  {author} {\bibfnamefont {V.}~\bibnamefont {Pomjakushin}}, \bibinfo {author}
  {\bibfnamefont {S.~W.}\ \bibnamefont {Lovesey}}, \bibinfo {author}
  {\bibfnamefont {A.}~\bibnamefont {Mirone}}, \bibinfo {author} {\bibfnamefont
  {J.}~\bibnamefont {Herrero-Martin}}, \bibinfo {author} {\bibfnamefont
  {C.}~\bibnamefont {Piamonteze}}, \ and\ \bibinfo {author} {\bibfnamefont
  {E.}~\bibnamefont {Pomjakushina}},\ }\href@noop {} {\bibfield  {journal}
  {\bibinfo  {journal} {Phys. Rev. Lett.},\ }\textbf {\bibinfo {volume}
  {103}},\ \bibinfo {eid} {097205} (\bibinfo {year} {2009})}\BibitemShut
  {NoStop}%
\bibitem [{\citenamefont {Akahoshi}\ \emph {et~al.}(2003)\citenamefont
  {Akahoshi}, \citenamefont {Uchida}, \citenamefont {Tomioka}, \citenamefont
  {Arima}, \citenamefont {Matsui},\ and\ \citenamefont
  {Tokura}}]{Akahoshi2003}%
  \BibitemOpen
  \bibfield  {author} {\bibinfo {author} {\bibfnamefont {D.}~\bibnamefont
  {Akahoshi}}, \bibinfo {author} {\bibfnamefont {M.}~\bibnamefont {Uchida}},
  \bibinfo {author} {\bibfnamefont {Y.}~\bibnamefont {Tomioka}}, \bibinfo
  {author} {\bibfnamefont {T.}~\bibnamefont {Arima}}, \bibinfo {author}
  {\bibfnamefont {Y.}~\bibnamefont {Matsui}}, \ and\ \bibinfo {author}
  {\bibfnamefont {Y.}~\bibnamefont {Tokura}},\ }\href@noop {} {\bibfield
  {journal} {\bibinfo  {journal} {Phys. Rev. Lett.},\ }\textbf {\bibinfo
  {volume} {90}},\ \bibinfo {pages} {177203} (\bibinfo {year}
  {2003})}\BibitemShut {NoStop}%
\bibitem [{\citenamefont {Akahoshi}\ \emph {et~al.}(2006)\citenamefont
  {Akahoshi}, \citenamefont {Uchida}, \citenamefont {Arima}, \citenamefont
  {Tomioka},\ and\ \citenamefont {Tokura}}]{Akahoshi2006}%
  \BibitemOpen
  \bibfield  {author} {\bibinfo {author} {\bibfnamefont {D.}~\bibnamefont
  {Akahoshi}}, \bibinfo {author} {\bibfnamefont {M.}~\bibnamefont {Uchida}},
  \bibinfo {author} {\bibfnamefont {T.}~\bibnamefont {Arima}}, \bibinfo
  {author} {\bibfnamefont {Y.}~\bibnamefont {Tomioka}}, \ and\ \bibinfo
  {author} {\bibfnamefont {Y.}~\bibnamefont {Tokura}},\ }\href@noop {}
  {\bibfield  {journal} {\bibinfo  {journal} {Phys. Rev. B},\ }\textbf
  {\bibinfo {volume} {74}},\ \bibinfo {pages} {012402} (\bibinfo {year}
  {2006})}\BibitemShut {NoStop}%
\bibitem [{\citenamefont {von Zimmermann}\ \emph {et~al.}(1999)\citenamefont
  {von Zimmermann}, \citenamefont {Hill}, \citenamefont {Gibbs}, \citenamefont
  {Blume}, \citenamefont {Casa}, \citenamefont {Keimer}, \citenamefont
  {Murakami}, \citenamefont {Tomioka},\ and\ \citenamefont
  {Tokura}}]{Zimmermann1999}%
  \BibitemOpen
  \bibfield  {author} {\bibinfo {author} {\bibfnamefont {M.}~\bibnamefont {von
  Zimmermann}}, \bibinfo {author} {\bibfnamefont {J.~P.}\ \bibnamefont {Hill}},
  \bibinfo {author} {\bibfnamefont {D.}~\bibnamefont {Gibbs}}, \bibinfo
  {author} {\bibfnamefont {M.}~\bibnamefont {Blume}}, \bibinfo {author}
  {\bibfnamefont {D.}~\bibnamefont {Casa}}, \bibinfo {author} {\bibfnamefont
  {B.}~\bibnamefont {Keimer}}, \bibinfo {author} {\bibfnamefont
  {Y.}~\bibnamefont {Murakami}}, \bibinfo {author} {\bibfnamefont
  {Y.}~\bibnamefont {Tomioka}}, \ and\ \bibinfo {author} {\bibfnamefont
  {Y.}~\bibnamefont {Tokura}},\ }\href@noop {} {\bibfield  {journal} {\bibinfo
  {journal} {Phys. Rev. Lett.},\ }\textbf {\bibinfo {volume} {83}},\ \bibinfo
  {pages} {4872} (\bibinfo {year} {1999})}\BibitemShut {NoStop}%
\bibitem [{\citenamefont {Garcia-Fernandez}\ \emph {et~al.}(2008)\citenamefont
  {Garcia-Fernandez}, \citenamefont {Staub}, \citenamefont {Bodenthin},
  \citenamefont {Lawrence}, \citenamefont {Mulders}, \citenamefont {Buckley},
  \citenamefont {Weyeneth}, \citenamefont {Pomjakushina},\ and\ \citenamefont
  {Conder}}]{Garcia-Fernandez-Sm-2008}%
  \BibitemOpen
  \bibfield  {author} {\bibinfo {author} {\bibfnamefont {M.}~\bibnamefont
  {Garcia-Fernandez}}, \bibinfo {author} {\bibfnamefont {U.}~\bibnamefont
  {Staub}}, \bibinfo {author} {\bibfnamefont {Y.}~\bibnamefont {Bodenthin}},
  \bibinfo {author} {\bibfnamefont {S.~M.}\ \bibnamefont {Lawrence}}, \bibinfo
  {author} {\bibfnamefont {A.~M.}\ \bibnamefont {Mulders}}, \bibinfo {author}
  {\bibfnamefont {C.~E.}\ \bibnamefont {Buckley}}, \bibinfo {author}
  {\bibfnamefont {S.}~\bibnamefont {Weyeneth}}, \bibinfo {author}
  {\bibfnamefont {E.}~\bibnamefont {Pomjakushina}}, \ and\ \bibinfo {author}
  {\bibfnamefont {K.}~\bibnamefont {Conder}},\ }\href@noop {} {\bibfield
  {journal} {\bibinfo  {journal} {Phys. Rev. B},\ }\textbf {\bibinfo {volume}
  {77}},\ \bibinfo {pages} {060402(R)} (\bibinfo {year} {2008})}\BibitemShut
  {NoStop}%
\bibitem [{\citenamefont {Thomas}\ \emph {et~al.}(2004)\citenamefont {Thomas},
  \citenamefont {Hill}, \citenamefont {Grenier}, \citenamefont {Kim},
  \citenamefont {Abbamonte}, \citenamefont {Venema}, \citenamefont {Rusydi},
  \citenamefont {Tomioka}, \citenamefont {Tokura}, \citenamefont {McMorrow},
  \citenamefont {Sawatzky},\ and\ \citenamefont {van Veenendaal}}]{Thomas2004}%
  \BibitemOpen
  \bibfield  {author} {\bibinfo {author} {\bibfnamefont {K.~J.}\ \bibnamefont
  {Thomas}}, \bibinfo {author} {\bibfnamefont {J.~P.}\ \bibnamefont {Hill}},
  \bibinfo {author} {\bibfnamefont {S.}~\bibnamefont {Grenier}}, \bibinfo
  {author} {\bibfnamefont {Y.~J.}\ \bibnamefont {Kim}}, \bibinfo {author}
  {\bibfnamefont {P.}~\bibnamefont {Abbamonte}}, \bibinfo {author}
  {\bibfnamefont {L.}~\bibnamefont {Venema}}, \bibinfo {author} {\bibfnamefont
  {A.}~\bibnamefont {Rusydi}}, \bibinfo {author} {\bibfnamefont
  {Y.}~\bibnamefont {Tomioka}}, \bibinfo {author} {\bibfnamefont
  {Y.}~\bibnamefont {Tokura}}, \bibinfo {author} {\bibfnamefont {D.~F.}\
  \bibnamefont {McMorrow}}, \bibinfo {author} {\bibfnamefont {G.}~\bibnamefont
  {Sawatzky}}, \ and\ \bibinfo {author} {\bibfnamefont {M.}~\bibnamefont {van
  Veenendaal}},\ }\href@noop {} {\bibfield  {journal} {\bibinfo  {journal}
  {Phys. Rev. Lett.},\ }\textbf {\bibinfo {volume} {92}},\ \bibinfo {pages}
  {237204} (\bibinfo {year} {2004})}\BibitemShut {NoStop}%
\bibitem [{\citenamefont {Staub}\ \emph {et~al.}(2005)\citenamefont {Staub},
  \citenamefont {Scagnoli}, \citenamefont {Mulders}, \citenamefont {Katsumata},
  \citenamefont {Honda}, \citenamefont {Grimmer}, \citenamefont {Horisberger},\
  and\ \citenamefont {Tonnerre}}]{Staub2005}%
  \BibitemOpen
  \bibfield  {author} {\bibinfo {author} {\bibfnamefont {U.}~\bibnamefont
  {Staub}}, \bibinfo {author} {\bibfnamefont {V.}~\bibnamefont {Scagnoli}},
  \bibinfo {author} {\bibfnamefont {A.~M.}\ \bibnamefont {Mulders}}, \bibinfo
  {author} {\bibfnamefont {K.}~\bibnamefont {Katsumata}}, \bibinfo {author}
  {\bibfnamefont {Z.}~\bibnamefont {Honda}}, \bibinfo {author} {\bibfnamefont
  {H.}~\bibnamefont {Grimmer}}, \bibinfo {author} {\bibfnamefont
  {M.}~\bibnamefont {Horisberger}}, \ and\ \bibinfo {author} {\bibfnamefont
  {J.~M.}\ \bibnamefont {Tonnerre}},\ }\href@noop {} {\bibfield  {journal}
  {\bibinfo  {journal} {Phys. Rev. B},\ }\textbf {\bibinfo {volume} {71}},\
  \bibinfo {pages} {214421} (\bibinfo {year} {2005})}\BibitemShut {NoStop}%
\bibitem [{\citenamefont {Dhesi}\ \emph {et~al.}(2004)\citenamefont {Dhesi},
  \citenamefont {Mirone}, \citenamefont {De~Nadai}, \citenamefont {Ohresser},
  \citenamefont {Bencok}, \citenamefont {Brookes}, \citenamefont {Reutler},
  \citenamefont {Revcolevschi}, \citenamefont {Tagliaferri}, \citenamefont
  {Toulemonde},\ and\ \citenamefont {van~der Laan}}]{Dhesi2004}%
  \BibitemOpen
  \bibfield  {author} {\bibinfo {author} {\bibfnamefont {S.~S.}\ \bibnamefont
  {Dhesi}}, \bibinfo {author} {\bibfnamefont {A.}~\bibnamefont {Mirone}},
  \bibinfo {author} {\bibfnamefont {C.}~\bibnamefont {De~Nadai}}, \bibinfo
  {author} {\bibfnamefont {P.}~\bibnamefont {Ohresser}}, \bibinfo {author}
  {\bibfnamefont {P.}~\bibnamefont {Bencok}}, \bibinfo {author} {\bibfnamefont
  {N.~B.}\ \bibnamefont {Brookes}}, \bibinfo {author} {\bibfnamefont
  {P.}~\bibnamefont {Reutler}}, \bibinfo {author} {\bibfnamefont
  {A.}~\bibnamefont {Revcolevschi}}, \bibinfo {author} {\bibfnamefont
  {A.}~\bibnamefont {Tagliaferri}}, \bibinfo {author} {\bibfnamefont
  {O.}~\bibnamefont {Toulemonde}}, \ and\ \bibinfo {author} {\bibfnamefont
  {G.}~\bibnamefont {van~der Laan}},\ }\href@noop {} {\bibfield  {journal}
  {\bibinfo  {journal} {Phys. Rev. Lett.},\ }\textbf {\bibinfo {volume} {92}},\
  \bibinfo {pages} {056403} (\bibinfo {year} {2004})}\BibitemShut {NoStop}%
\bibitem [{\citenamefont {Wilkins}\ \emph {et~al.}(2003)\citenamefont
  {Wilkins}, \citenamefont {Spencer}, \citenamefont {Hatton}, \citenamefont
  {Collins}, \citenamefont {Roper}, \citenamefont {Prabhakaran},\ and\
  \citenamefont {Boothroyd}}]{Wilkins2003}%
  \BibitemOpen
  \bibfield  {author} {\bibinfo {author} {\bibfnamefont {S.~B.}\ \bibnamefont
  {Wilkins}}, \bibinfo {author} {\bibfnamefont {P.~D.}\ \bibnamefont
  {Spencer}}, \bibinfo {author} {\bibfnamefont {P.~D.}\ \bibnamefont {Hatton}},
  \bibinfo {author} {\bibfnamefont {S.~P.}\ \bibnamefont {Collins}}, \bibinfo
  {author} {\bibfnamefont {M.~D.}\ \bibnamefont {Roper}}, \bibinfo {author}
  {\bibfnamefont {D.}~\bibnamefont {Prabhakaran}}, \ and\ \bibinfo {author}
  {\bibfnamefont {A.~T.}\ \bibnamefont {Boothroyd}},\ }\href@noop {} {\bibfield
   {journal} {\bibinfo  {journal} {Phys. Rev. Lett.},\ }\textbf {\bibinfo
  {volume} {91}},\ \bibinfo {pages} {167205} (\bibinfo {year}
  {2003})}\BibitemShut {NoStop}%
\bibitem [{\citenamefont {Beale}\ \emph {et~al.}(2009)\citenamefont {Beale},
  \citenamefont {Bland}, \citenamefont {Johnson}, \citenamefont {Hatton},
  \citenamefont {Cezar}, \citenamefont {Dhesi}, \citenamefont {Zimmermann},
  \citenamefont {Prabhakaran},\ and\ \citenamefont {Boothroyd}}]{Beale2009}%
  \BibitemOpen
  \bibfield  {author} {\bibinfo {author} {\bibfnamefont {T.~A.~W.}\
  \bibnamefont {Beale}}, \bibinfo {author} {\bibfnamefont {S.~R.}\ \bibnamefont
  {Bland}}, \bibinfo {author} {\bibfnamefont {R.~D.}\ \bibnamefont {Johnson}},
  \bibinfo {author} {\bibfnamefont {P.~D.}\ \bibnamefont {Hatton}}, \bibinfo
  {author} {\bibfnamefont {J.~C.}\ \bibnamefont {Cezar}}, \bibinfo {author}
  {\bibfnamefont {S.~S.}\ \bibnamefont {Dhesi}}, \bibinfo {author}
  {\bibfnamefont {M.~V.}\ \bibnamefont {Zimmermann}}, \bibinfo {author}
  {\bibfnamefont {D.}~\bibnamefont {Prabhakaran}}, \ and\ \bibinfo {author}
  {\bibfnamefont {A.~T.}\ \bibnamefont {Boothroyd}},\ }\href@noop {} {\bibfield
   {journal} {\bibinfo  {journal} {Phys. Rev. B},\ }\textbf {\bibinfo {volume}
  {79}},\ \bibinfo {pages} {054433} (\bibinfo {year} {2009})}\BibitemShut
  {NoStop}%
\bibitem [{\citenamefont {Staub}\ \emph {et~al.}(2009)\citenamefont {Staub},
  \citenamefont {Garcia-Fernandez}, \citenamefont {Bodenthin}, \citenamefont
  {Scagnoli}, \citenamefont {De~Souza}, \citenamefont {Garganourakis},
  \citenamefont {Pomjakushina},\ and\ \citenamefont {Conder}}]{Staub2009}%
  \BibitemOpen
  \bibfield  {author} {\bibinfo {author} {\bibfnamefont {U.}~\bibnamefont
  {Staub}}, \bibinfo {author} {\bibfnamefont {M.}~\bibnamefont
  {Garcia-Fernandez}}, \bibinfo {author} {\bibfnamefont {Y.}~\bibnamefont
  {Bodenthin}}, \bibinfo {author} {\bibfnamefont {V.}~\bibnamefont {Scagnoli}},
  \bibinfo {author} {\bibfnamefont {R.~A.}\ \bibnamefont {De~Souza}}, \bibinfo
  {author} {\bibfnamefont {M.}~\bibnamefont {Garganourakis}}, \bibinfo {author}
  {\bibfnamefont {E.}~\bibnamefont {Pomjakushina}}, \ and\ \bibinfo {author}
  {\bibfnamefont {K.}~\bibnamefont {Conder}},\ }\href@noop {} {\bibfield
  {journal} {\bibinfo  {journal} {Phys. Rev. B},\ }\textbf {\bibinfo {volume}
  {79}},\ \bibinfo {pages} {224419} (\bibinfo {year} {2009})}\BibitemShut
  {NoStop}%
\bibitem [{\citenamefont {Fischer}\ \emph {et~al.}(2000)\citenamefont
  {Fischer}, \citenamefont {Frey}, \citenamefont {Koch}, \citenamefont
  {Konnecke}, \citenamefont {Pomjakushin}, \citenamefont {Schefer},
  \citenamefont {Thut}, \citenamefont {Schlumpf}, \citenamefont {Burge},
  \citenamefont {Greuter}, \citenamefont {Bondt},\ and\ \citenamefont
  {Berruyer}}]{Fischer_2000}%
  \BibitemOpen
  \bibfield  {author} {\bibinfo {author} {\bibfnamefont {P.}~\bibnamefont
  {Fischer}}, \bibinfo {author} {\bibfnamefont {G.}~\bibnamefont {Frey}},
  \bibinfo {author} {\bibfnamefont {M.}~\bibnamefont {Koch}}, \bibinfo {author}
  {\bibfnamefont {M.}~\bibnamefont {Konnecke}}, \bibinfo {author}
  {\bibfnamefont {V.}~\bibnamefont {Pomjakushin}}, \bibinfo {author}
  {\bibfnamefont {J.}~\bibnamefont {Schefer}}, \bibinfo {author} {\bibfnamefont
  {R.}~\bibnamefont {Thut}}, \bibinfo {author} {\bibfnamefont {N.}~\bibnamefont
  {Schlumpf}}, \bibinfo {author} {\bibfnamefont {R.}~\bibnamefont {Burge}},
  \bibinfo {author} {\bibfnamefont {U.}~\bibnamefont {Greuter}}, \bibinfo
  {author} {\bibfnamefont {S.}~\bibnamefont {Bondt}}, \ and\ \bibinfo {author}
  {\bibfnamefont {E.}~\bibnamefont {Berruyer}},\ }\href@noop {} {\bibfield
  {journal} {\bibinfo  {journal} {Physica B},\ }\textbf {\bibinfo {volume}
  {276}},\ \bibinfo {pages} {146} (\bibinfo {year} {2000})}\BibitemShut
  {NoStop}%
\bibitem [{\citenamefont {Rodríguez-Carvajal}(1993)}]{Rodriguez-Carvajal1993}%
  \BibitemOpen
  \bibfield  {author} {\bibinfo {author} {\bibfnamefont {J.}~\bibnamefont
  {Rodríguez-Carvajal}},\ }\href@noop {} {\bibfield  {journal} {\bibinfo
  {journal} {Physica B},\ }\textbf {\bibinfo {volume} {192}},\ \bibinfo {pages}
  {55} (\bibinfo {year} {1993})}\BibitemShut {NoStop}%
\bibitem [{\citenamefont {Staub}\ \emph {et~al.}(2008)\citenamefont {Staub},
  \citenamefont {Scagnoli}, \citenamefont {Bodenthin}, \citenamefont
  {Garcia-Fernandez}, \citenamefont {Wetter}, \citenamefont {Mulders},
  \citenamefont {Grimmer},\ and\ \citenamefont {Horisberger}}]{Staub2008}%
  \BibitemOpen
  \bibfield  {author} {\bibinfo {author} {\bibfnamefont {U.}~\bibnamefont
  {Staub}}, \bibinfo {author} {\bibfnamefont {V.}~\bibnamefont {Scagnoli}},
  \bibinfo {author} {\bibfnamefont {Y.}~\bibnamefont {Bodenthin}}, \bibinfo
  {author} {\bibfnamefont {M.}~\bibnamefont {Garcia-Fernandez}}, \bibinfo
  {author} {\bibfnamefont {R.}~\bibnamefont {Wetter}}, \bibinfo {author}
  {\bibfnamefont {A.~M.}\ \bibnamefont {Mulders}}, \bibinfo {author}
  {\bibfnamefont {H.}~\bibnamefont {Grimmer}}, \ and\ \bibinfo {author}
  {\bibfnamefont {M.}~\bibnamefont {Horisberger}},\ }\href@noop {} {\bibfield
  {journal} {\bibinfo  {journal} {J. Synchrotron Rad.},\ }\textbf {\bibinfo
  {volume} {15}},\ \bibinfo {pages} {469} (\bibinfo {year} {2008})}\BibitemShut
  {NoStop}%
\bibitem [{\citenamefont {Radaelli}\ \emph {et~al.}(1997)\citenamefont
  {Radaelli}, \citenamefont {Cox}, \citenamefont {Marezio},\ and\ \citenamefont
  {Cheong}}]{Radaelli1997}%
  \BibitemOpen
  \bibfield  {author} {\bibinfo {author} {\bibfnamefont {P.~G.}\ \bibnamefont
  {Radaelli}}, \bibinfo {author} {\bibfnamefont {D.~E.}\ \bibnamefont {Cox}},
  \bibinfo {author} {\bibfnamefont {M.}~\bibnamefont {Marezio}}, \ and\
  \bibinfo {author} {\bibfnamefont {S.-W.}\ \bibnamefont {Cheong}},\
  }\href@noop {} {\bibfield  {journal} {\bibinfo  {journal} {Phys. Rev. B},\
  }\textbf {\bibinfo {volume} {55}},\ \bibinfo {pages} {3015} (\bibinfo {year}
  {1997})}\BibitemShut {NoStop}%
\bibitem [{\citenamefont {Goff}\ and\ \citenamefont
  {Attfield}(2004)}]{Goff2004}%
  \BibitemOpen
  \bibfield  {author} {\bibinfo {author} {\bibfnamefont {R.~J.}\ \bibnamefont
  {Goff}}\ and\ \bibinfo {author} {\bibfnamefont {J.~P.}\ \bibnamefont
  {Attfield}},\ }\href@noop {} {\bibfield  {journal} {\bibinfo  {journal}
  {Phys. Rev. B},\ }\textbf {\bibinfo {volume} {70}},\ \bibinfo {pages}
  {140404} (\bibinfo {year} {2004})}\BibitemShut {NoStop}%
\bibitem [{\citenamefont {Blasco}\ \emph {et~al.}(1997)\citenamefont {Blasco},
  \citenamefont {Garcia}, \citenamefont {deTeresa}, \citenamefont {Ibarra},
  \citenamefont {Perez}, \citenamefont {Algarabel}, \citenamefont {Marquina},\
  and\ \citenamefont {Ritter}}]{Blasco1997}%
  \BibitemOpen
  \bibfield  {author} {\bibinfo {author} {\bibfnamefont {J.}~\bibnamefont
  {Blasco}}, \bibinfo {author} {\bibfnamefont {J.}~\bibnamefont {Garcia}},
  \bibinfo {author} {\bibfnamefont {J.~M.}\ \bibnamefont {deTeresa}}, \bibinfo
  {author} {\bibfnamefont {M.~R.}\ \bibnamefont {Ibarra}}, \bibinfo {author}
  {\bibfnamefont {J.}~\bibnamefont {Perez}}, \bibinfo {author} {\bibfnamefont
  {P.~A.}\ \bibnamefont {Algarabel}}, \bibinfo {author} {\bibfnamefont
  {C.}~\bibnamefont {Marquina}}, \ and\ \bibinfo {author} {\bibfnamefont
  {C.}~\bibnamefont {Ritter}},\ }\href@noop {} {\bibfield  {journal} {\bibinfo
  {journal} {J. Phy.: Condens. Matt.},\ }\textbf {\bibinfo {volume} {9}},\
  \bibinfo {pages} {10321} (\bibinfo {year} {1997})}\BibitemShut {NoStop}%
\bibitem [{\citenamefont {Woodward}\ \emph {et~al.}(1999)\citenamefont
  {Woodward}, \citenamefont {Cox}, \citenamefont {Vogt}, \citenamefont {Rao},\
  and\ \citenamefont {Cheetham}}]{Woodward1999}%
  \BibitemOpen
  \bibfield  {author} {\bibinfo {author} {\bibfnamefont {P.~M.}\ \bibnamefont
  {Woodward}}, \bibinfo {author} {\bibfnamefont {D.~E.}\ \bibnamefont {Cox}},
  \bibinfo {author} {\bibfnamefont {T.}~\bibnamefont {Vogt}}, \bibinfo {author}
  {\bibfnamefont {C.~N.~R.}\ \bibnamefont {Rao}}, \ and\ \bibinfo {author}
  {\bibfnamefont {A.~K.}\ \bibnamefont {Cheetham}},\ }\href@noop {} {\bibfield
  {journal} {\bibinfo  {journal} {Chem. Of Mat.},\ }\textbf {\bibinfo {volume}
  {11}},\ \bibinfo {pages} {3528} (\bibinfo {year} {1999})}\BibitemShut
  {NoStop}%
\bibitem [{\citenamefont {Richard}\ \emph {et~al.}(1999)\citenamefont
  {Richard}, \citenamefont {Schuddinck}, \citenamefont {Van~Tendeloo},
  \citenamefont {Millange}, \citenamefont {Hervieu}, \citenamefont
  {Caignaert},\ and\ \citenamefont {Raveau}}]{Richard1999}%
  \BibitemOpen
  \bibfield  {author} {\bibinfo {author} {\bibfnamefont {O.}~\bibnamefont
  {Richard}}, \bibinfo {author} {\bibfnamefont {W.}~\bibnamefont {Schuddinck}},
  \bibinfo {author} {\bibfnamefont {G.}~\bibnamefont {Van~Tendeloo}}, \bibinfo
  {author} {\bibfnamefont {F.}~\bibnamefont {Millange}}, \bibinfo {author}
  {\bibfnamefont {M.}~\bibnamefont {Hervieu}}, \bibinfo {author} {\bibfnamefont
  {V.}~\bibnamefont {Caignaert}}, \ and\ \bibinfo {author} {\bibfnamefont
  {B.}~\bibnamefont {Raveau}},\ }\href@noop {} {\bibfield  {journal} {\bibinfo
  {journal} {Act. Cryst. Sect. A},\ }\textbf {\bibinfo {volume} {55}},\
  \bibinfo {pages} {704} (\bibinfo {year} {1999})}\BibitemShut {NoStop}%
\bibitem [{\citenamefont {Williams}\ \emph {et~al.}(2005)\citenamefont
  {Williams}, \citenamefont {Attfield},\ and\ \citenamefont
  {Redfern}}]{Williams2005}%
  \BibitemOpen
  \bibfield  {author} {\bibinfo {author} {\bibfnamefont {A.~J.}\ \bibnamefont
  {Williams}}, \bibinfo {author} {\bibfnamefont {J.~P.}\ \bibnamefont
  {Attfield}}, \ and\ \bibinfo {author} {\bibfnamefont {S.~A.~T.}\ \bibnamefont
  {Redfern}},\ }\href@noop {} {\bibfield  {journal} {\bibinfo  {journal} {Phys.
  Rev. B},\ }\textbf {\bibinfo {volume} {72}},\ \bibinfo {pages} {184426}
  (\bibinfo {year} {2005})}\BibitemShut {NoStop}%
\bibitem [{\citenamefont {Pomjakushina}\ \emph {et~al.}(2006)\citenamefont
  {Pomjakushina}, \citenamefont {Conder},\ and\ \citenamefont
  {Pomjakushin}}]{key-69}%
  \BibitemOpen
  \bibfield  {author} {\bibinfo {author} {\bibfnamefont {E.}~\bibnamefont
  {Pomjakushina}}, \bibinfo {author} {\bibfnamefont {K.}~\bibnamefont
  {Conder}}, \ and\ \bibinfo {author} {\bibfnamefont {V.}~\bibnamefont
  {Pomjakushin}},\ }\href@noop {} {\bibfield  {journal} {\bibinfo  {journal}
  {Phys. Rev. B},\ }\textbf {\bibinfo {volume} {73}},\ \bibinfo {pages}
  {113105} (\bibinfo {year} {2006})}\BibitemShut {NoStop}%
\bibitem [{\citenamefont {Staub}\ \emph {et~al.}(2006)\citenamefont {Staub},
  \citenamefont {Scagnoli}, \citenamefont {Mulders}, \citenamefont {Janousch},
  \citenamefont {Honda},\ and\ \citenamefont {Tonnerre}}]{Staub2006}%
  \BibitemOpen
  \bibfield  {author} {\bibinfo {author} {\bibfnamefont {U.}~\bibnamefont
  {Staub}}, \bibinfo {author} {\bibfnamefont {V.}~\bibnamefont {Scagnoli}},
  \bibinfo {author} {\bibfnamefont {A.~M.}\ \bibnamefont {Mulders}}, \bibinfo
  {author} {\bibfnamefont {M.}~\bibnamefont {Janousch}}, \bibinfo {author}
  {\bibfnamefont {Z.}~\bibnamefont {Honda}}, \ and\ \bibinfo {author}
  {\bibfnamefont {J.~M.}\ \bibnamefont {Tonnerre}},\ }\href@noop {} {\bibfield
  {journal} {\bibinfo  {journal} {Europhys. Lett.},\ }\textbf {\bibinfo
  {volume} {76}},\ \bibinfo {pages} {926} (\bibinfo {year} {2006})}\BibitemShut
  {NoStop}%
\bibitem [{\citenamefont {Garcia Mu\~{n}oz}\ \emph {et~al.}(1996)\citenamefont
  {Garcia Mu\~{n}oz}, \citenamefont {Fontcuberta}, \citenamefont {Suaaidi},\
  and\ \citenamefont {Obradors}}]{Garcia-Munoz1996}%
  \BibitemOpen
  \bibfield  {author} {\bibinfo {author} {\bibfnamefont {J.~L.}\ \bibnamefont
  {Garcia Mu\~{n}oz}}, \bibinfo {author} {\bibfnamefont {J.}~\bibnamefont
  {Fontcuberta}}, \bibinfo {author} {\bibfnamefont {M.}~\bibnamefont
  {Suaaidi}}, \ and\ \bibinfo {author} {\bibfnamefont {X.}~\bibnamefont
  {Obradors}},\ }\href@noop {} {\bibfield  {journal} {\bibinfo  {journal} {J.
  Phys.: Condens. Matter},\ }\textbf {\bibinfo {volume} {8}},\ \bibinfo {pages}
  {L787} (\bibinfo {year} {1996})}\BibitemShut {NoStop}%
\bibitem [{\citenamefont {Cox}\ \emph {et~al.}(2008)\citenamefont {Cox},
  \citenamefont {Singleton}, \citenamefont {McDonald}, \citenamefont
  {Migliori},\ and\ \citenamefont {Littlewood}}]{Cox2008}%
  \BibitemOpen
  \bibfield  {author} {\bibinfo {author} {\bibfnamefont {S.}~\bibnamefont
  {Cox}}, \bibinfo {author} {\bibfnamefont {J.}~\bibnamefont {Singleton}},
  \bibinfo {author} {\bibfnamefont {R.~D.}\ \bibnamefont {McDonald}}, \bibinfo
  {author} {\bibfnamefont {A.}~\bibnamefont {Migliori}}, \ and\ \bibinfo
  {author} {\bibfnamefont {P.~B.}\ \bibnamefont {Littlewood}},\ }\href@noop {}
  {\bibfield  {journal} {\bibinfo  {journal} {Nat. Mater.},\ }\textbf {\bibinfo
  {volume} {7}},\ \bibinfo {pages} {25} (\bibinfo {year} {2008})},\ ISSN
  \bibinfo {issn} {1476-1122}\BibitemShut {NoStop}%
\bibitem [{\citenamefont {Mirone}\ \emph {et~al.}(2006)\citenamefont {Mirone},
  \citenamefont {Dhesi},\ and\ \citenamefont {van~der Laan}}]{Mirone2006}%
  \BibitemOpen
  \bibfield  {author} {\bibinfo {author} {\bibfnamefont {A.}~\bibnamefont
  {Mirone}}, \bibinfo {author} {\bibfnamefont {S.~S.}\ \bibnamefont {Dhesi}}, \
  and\ \bibinfo {author} {\bibfnamefont {G.}~\bibnamefont {van~der Laan}},\
  }\href@noop {} {\bibfield  {journal} {\bibinfo  {journal} {Eur. Phys. J. B},\
  }\textbf {\bibinfo {volume} {53}},\ \bibinfo {pages} {23} (\bibinfo {year}
  {2006})}\BibitemShut {NoStop}%
\bibitem [{\citenamefont {Herrero-Martin}\ \emph {et~al.}(2010)\citenamefont
  {Herrero-Martin}, \citenamefont {Mirone}, \citenamefont {Fernadez-Rodriguez},
  \citenamefont {Glatzel}, \citenamefont {Garcia}, \citenamefont {Blasco},\
  and\ \citenamefont {Geck}}]{Herrero-Martin2010}%
  \BibitemOpen
  \bibfield  {author} {\bibinfo {author} {\bibfnamefont {J.}~\bibnamefont
  {Herrero-Martin}}, \bibinfo {author} {\bibfnamefont {A.}~\bibnamefont
  {Mirone}}, \bibinfo {author} {\bibfnamefont {J.}~\bibnamefont
  {Fernadez-Rodriguez}}, \bibinfo {author} {\bibfnamefont {P.}~\bibnamefont
  {Glatzel}}, \bibinfo {author} {\bibfnamefont {J.}~\bibnamefont {Garcia}},
  \bibinfo {author} {\bibfnamefont {J.}~\bibnamefont {Blasco}}, \ and\ \bibinfo
  {author} {\bibfnamefont {J.}~\bibnamefont {Geck}},\ }\href@noop {} {\bibfield
   {journal} {\bibinfo  {journal} {Phys. Rev. B},\ }\textbf {\bibinfo {volume}
  {82}},\ \bibinfo {pages} {075112} (\bibinfo {year} {2010})}\BibitemShut
  {NoStop}%
\bibitem [{\citenamefont {Volja}\ \emph {et~al.}(2010)\citenamefont {Volja},
  \citenamefont {Yin},\ and\ \citenamefont {Ku}}]{Volja2010}%
  \BibitemOpen
  \bibfield  {author} {\bibinfo {author} {\bibfnamefont {D.}~\bibnamefont
  {Volja}}, \bibinfo {author} {\bibfnamefont {W.-G.}\ \bibnamefont {Yin}}, \
  and\ \bibinfo {author} {\bibfnamefont {W.}~\bibnamefont {Ku}},\ }\href@noop
  {} {\bibfield  {journal} {\bibinfo  {journal} {Europhys. Lett.},\ }\textbf
  {\bibinfo {volume} {89}},\ \bibinfo {pages} {27008} (\bibinfo {year}
  {2010})}\BibitemShut {NoStop}%
\bibitem [{\citenamefont {Stojic}\ \emph {et~al.}(2005)\citenamefont {Stojic},
  \citenamefont {Binggeli},\ and\ \citenamefont {Altarelli}}]{Stojic2005}%
  \BibitemOpen
  \bibfield  {author} {\bibinfo {author} {\bibfnamefont {N.}~\bibnamefont
  {Stojic}}, \bibinfo {author} {\bibfnamefont {N.}~\bibnamefont {Binggeli}}, \
  and\ \bibinfo {author} {\bibfnamefont {M.}~\bibnamefont {Altarelli}},\
  }\href@noop {} {\bibfield  {journal} {\bibinfo  {journal} {Phys. Rev. B},\
  }\textbf {\bibinfo {volume} {72}},\ \bibinfo {pages} {104108} (\bibinfo
  {year} {2005})}\BibitemShut {NoStop}%
\end{thebibliography}%

\end{document}